\documentclass[10pt,final,journal,twocolumn]{IEEEtran}
\usepackage{graphicx,cite,epsfig,amssymb,amsmath}

\begin{document}
\title{
{User Mobility Evaluation for 5G Small Cell Networks Based on Individual Mobility Model}}

\author{
 Xiaohu Ge$^1$, Junliang Ye$^1$, Yang Yang$^{2, 3}$, Qiang Li$^1$\\
\vspace{0.10cm}
{
$^1$School of Electronic Information and Communications\\
Huazhong University of Science and Technology, Wuhan 430074, Hubei, P. R.
China.\\}
\vspace{0.1cm}
$^2$Shanghai Institute of Microsystem and Information technology (SIMIT), \\
Chinese Academy of Sciences, Shanghai 200050.\\
$^3$ShanghaiTech University, China.\\
\vspace{0.1cm}
\vspace{-0.5cm}

\thanks{\scriptsize{Submitted to IEEE JSAC Series on Emerging Technologies.}}
\thanks{\scriptsize{Correspondence author: Dr. Qiang Li, Tel: +86 (0)27 87557942, Fax: +86 (0)27 87557943, Email: qli\_patrick@mail.hust.edu.cn.}}
\thanks{\scriptsize{The authors would like to acknowledge the support from the National Natural Science Foundation of China (NSFC) under the grants 61301128 and 61461136004, NFSC Major International Joint Research Project under the grant 61210002, the Ministry of Science and Technology (MOST) of China under the grants 2015FDG12580 and 2014DFA11640, the Fundamental Research Funds for the Central Universities under the grant 2015XJGH011 and 2014QN155, the Special Research Fund for the Doctoral Program of Higher Education (SRFDP) under grant 20130142120044. Yang's research is partially supported by the Science and Technology Commission of Shanghai Municipality (STCSM) under grant 15511103200, the National Natural Science Foundation of China (NSFC) under grants 61231009 and 61461136003, and the Ministry of Science and Technolgoy (MOST) under grant 2014DFE10160. This research is partially supported by the EU FP7-PEOPLE-IRSES, project acronym S2EuNet (grant no. 247083), project acronym WiNDOW (grant no. 318992) and project acronym CROWN (grant no. 610524).}}
}
\maketitle

\markboth{IEEE Journal on Selected Areas in Communications, Vol. XX, No. Y,
Month 2016} {Ge etc.: User Mobility Evaluation for 5G Small Cell Networks Based on Individual Mobility Model\ldots}%
\begin{abstract}

With small cell networks becoming core parts of the fifth generation (5G) cellular networks, it is an important problem
to evaluate the impact of user mobility on 5G small cell networks. However, the tendency and
clustering habits in human activities have not been considered in traditional user mobility models.
In this paper, human tendency and clustering behaviors are first considered to evaluate the user
mobility performance for 5G small cell networks based on individual mobility model (IMM). As key
contributions, user pause probability, user arrival and departure probabilities are derived in this paper for
evaluating the user mobility performance in a hotspot-type 5G small cell network.
Furthermore, coverage probabilities of small cell and macro cell BSs are derived for all users in 5G
small cell networks, respectively. Compared with the traditional random waypoint (RWP) model, IMM provides a different viewpoint to investigate the impact of human tendency and
clustering behaviors on the performance of 5G small cell networks.

\end{abstract}

\section{Introduction}
\label{sec1}

To satisfy the high transmission rate requirement in the fifth generation (5G) cellular networks, one of the effective approaches is to reduce the coverage radius of  cells, thus reducing the communication distances between users and base stations (BSs) \cite{Demestichas13}. Moreover, two of 5G key technologies, i.e., massive multiple-input multi-output (MIMO) antenna and millimeter wave wireless transmission technologies further reduce the wireless transmission distance in 5G wireless communication systems \cite{Xiaoyu15} \cite{Jeffrey14}. Hence, 5G cellular network is a type of small cell networks. On the other hand, the impact of user mobility on 5G small cell networks is enlarged with the decrease of the cell coverage radius \cite{Fabio15,Hwang13,Vasudevan13}. Therefore, it is an important problem to evaluate the impact of the user mobility on 5G small cell networks considering human activity habits.

Knowledge of human mobility behavior is essential for all investigations in which the relative locations of mobility users are important. By measuring the entropy of each individual¡¯s trajectory, Song {\em et al.} found a 93\% potential predictability in user mobility across the whole user base  \cite{Chaoming10}. After this, human mobility models have attracted more attentions from telecommunication researchers and some human mobility models have been used for wireless networks, such as cellular networks and vehicular networks. In conventional mobile communication investigations, user mobility models have been classified four types, i.e., the random walk model, the random waypoint (RWP) model, the fluid flow model and Gauss-Markov model  \cite{Hong99,Camp02,Akyildiz00,Bettstetter04,Pantelis14,Yang12,Wang14,Chung12,Liang03,Wang12,Yeow07}. The random walk model is originally used for describing the atom mobility in physics and chemistry topics. The mobility direction and velocity of random walk model are fully stochastic and independent with the past status \cite{Camp02}. A new approach was proposed to simplify the two-dimensional random walk models capturing the movement of mobile users in personal communications services networks \cite{Akyildiz00}. By adding the pause time into the random walk model, the random waypoint model was presented to simulate the mobility user in wireless networks \cite{Bettstetter04}.  Based on the random waypoint model, the performance bounds and benefits of wireless networks was evaluated for a user centric solution \cite{Pantelis14}. A position-based opportunistic routing protocol was developed to reduce the latency incurred by local route recovery in ad hoc networks where the mobile user follows the random waypoint model \cite{Yang12}. The fluid flow model is widely used for calculating the probability that a mobile node accesses a closed area, such as the cellular cell \cite{Wang14}. Utilizing the fluid flow model, a location area residence time was analyzed for the mobile user in cellular networks \cite{Chung12}. The memory of mobile user is considered in the Gauss-Markov model, where the next location of mobile user depends on the last location of mobile user  \cite{Liang03}. Accounting for the Gauss-Markov model, an analytical framework was proposed to evaluate the cost of mobility management for wireless personal communication service networks \cite{Wang12}.Adopting the target trajectory model from the Gauss-Markov mobility model in wireless sensor networks, a target-tracking algorithm was developed to conserve the energy by reducing the rate of sensing in temporal management \cite{Yeow07}.Except of above four mobility models, some user mobility models measured from real data have been adopted for wireless networks \cite{Wu11,Lin13}. By mining the extensive trace datasets of vehicles in an urban environment through conditional entropy analysis, a packet delivery probability was derived for vehicular networks \cite{Wu11}. Based on measured results from cellular networks, e.g., handoff rates and call arrival rates, a solution was proposed to predict how people spread from one location to another after a period of time \cite{Lin13}. However, the tendency and clustering habits in human activities have not been considered for above human mobility models. In real scenarios, people habitually stay at some locations for a long time compared with other locations. To describe the user clustering habit, the concept of community was introduced to evaluate the user mobility in wireless networks \cite{Hsu09}. Based on the human mobility trajectory measured from real data, the individual mobility model considering the human mobility tendency habit was proposed to describe the human mobility in real world \cite{Song10}.

The impact of the user mobility on small cell networks has been studied in \cite{Daniel15,Prasad13,Prasad15,Pedersen13,Sanguinetti15}. By modeling positions of mobile users as an independent Poisson point process in each time slot, the backhaul delay of heterogeneous networks including small cell networks and macro cell networks was analyzed in \cite{Daniel15}. A new small cell discovery mechanism based on the user mobility state was investigated to improve the energy efficiency and access efficiency of LTE-Advanced heterogeneous wireless networks \cite{Prasad13}. Considering mobile user scenarios, an energy efficient and backhaul aware small cell activation mechanism with the use of dual connectivity was evaluated for 5G small cell networks \cite{Prasad15}. To reduce the signal overhead, a new scheme was proposed by having the mobile users autonomously decide small cell addition, remove, and change without any explicit signaling of measurement events to small cell networks or any signaling of handoff commends from small cell networks \cite{Pedersen13}. Based on 5G wireless network scenarios, the downlink and uplink of power consumption was evaluated for small cell networks where static and low mobility users are associated with macro cell BSs and high mobility user are associated with small cell BSs \cite{Sanguinetti15}.

However, the user mobility performance of small cell networks in all aforementioned research considers only simple mobility models, such as the random walk model or the random waypoint model. Besides, the impact of human tendency and clustering behaviors on small cell networks has not been investigated. Moreover, detailed investigation of IMM used for small cell networks is surprisingly rare in the open literature. Motived by above gaps, we first analyze the user mobility performance and derive coverage probabilities for 5G small cell networks based on IMM. The contributions and novelties of this paper are summarized as follows.

\begin{enumerate}
\item Based on IMM, the user arrival, departure and pause probabilities are derived to evaluate the user mobility performance for 5G small cell networks with the community area.
\item Based on the proposed arrival and departure probabilities, the coverage probabilities inside and outside the community are derived for 5G small cell networks. Considering the group cell scheme, the number of available small cell BSs is derived for the static user in 5G small cell networks.
\item Furthermore, the coverage probability of macro cell BSs is derived for the mobile user in 5G small cell networks. Numerical simulations validate the proposed coverage probabilities for 5G small cell networks.

\end{enumerate}

The rest of this paper is organized as follows. Section II describes the system model of 5G small cell networks  within a community area and the IMM. In Section III, the user arrival, departure and pause probabilities are derived to evaluate the user mobility performance for 5G small cell networks with community area. Furthermore, in Section IV the coverage probabilities inside and outside the community are derived for 5G small cell networks. Considering the group cell scheme, the number of available small cell BSs is derived for the static user in 5G small cell networks. Moreover, the coverage probability of macro cell BSs is derived for the mobile user in 5G small cell networks. In Section V, Numerical results validate the proposed coverage probabilities for 5G small cell networks. Finally, Section VI concludes this paper.

\section{System Model}
\label{sec2}
\subsection{Network Model}
Assume that both users and BSs are located in a finite plane $\mathbb{R}_t^2$ whose area is ${S_t}$. Moreover, a rectangle $\mathbb{R}_c^2$ is located in the finite plane $\mathbb{R}_t^2$ and the area of rectangle is ${S_c}$. The complement region of rectangle $\mathbb{R}_c^2$ is denoted as $\mathbb{R}_s^2$ whose area is ${S_s}$. In this paper, the concept of community is introduced to represents a specified region where a large number of users are assembled \cite{Hsu09}. Without loss of generality, the rectangle $\mathbb{R}_c^2$ is configured as a community in the finite plane $\mathbb{R}_t^2$. Macro cell BSs and small cell BSs are located in the finite plane $\mathbb{R}_t^2$. Macro cell BSs are assumed to be governed by a Poisson point process distribution with the density ${\lambda _{m,{\text{BS}}}}$.The macro cell boundary, which can be obtained through the Delaunay Triangulation method by connecting the perpendicular bisector lines between each pair of macro cell BSs, splits the plane $\mathbb{R}_t^2$. into irregular polygons that correspond to different cell coverage areas. This stochastic and irregular topology forms a so-called Poisson-Voronoi tessellation (PVT) \cite{Baccelli97,Ge15}. Small cell BSs inside and outside the community are assumed to follow uniform distributions with densities ${\lambda _{c,{\text{BS}}}}$ and ${\lambda _{s,{\text{BS}}}}$, respectively. Macro cells and small cells are overlapped in the infinite plane $\mathbb{R}_t^2$. Macro cell BSs and small cell BSs transmit wireless signal in different frequencies. Hence, there does not exist the interference between macro cell networks and small cell networks. Users inside and outside the community are assumed to follow uniform distributions with densities ${\lambda _s}$ and ${\lambda _c}$, respectively. Based on the community concept, the following constraints are configured: ${\lambda _c} > {\lambda _s}$ and ${\lambda _{c,BS}} > {\lambda _{s,BS}}$. An illustration of users and BSs deployment is depicted in Fig. 1.

When the orthogonal frequency division multiplexing (OFDM) scheme is adopted by BSs to support multi-user transmission in this paper, the co-channel interference generated from the intra-cell is ignored. Only downlink transmission is studied in this paper. Moreover, the static user is assumed to be associated with small cell BSs and the mobile user is assumed to be associated with macro cell BSs \cite{Sanam11}. To realize the high transmission rate, the group cell scheme is proposed for the static user in small cell networks. In the group cell scheme, the static user can be associated with multiply small cell BSs if signal-to-interference-plus-noise ratios (SINRs) over wireless links are larger than or equal to a given threshold ${\gamma _0}$. To simplify derivation, the static user is only associated with small cell BSs inside the community when the static user is located inside the community. Similarly, the static user is only associated with small cell BSs outside the community when the static user is located outside the community.
\begin{figure}
\vspace{0.1in}
\centerline{\includegraphics[width=8cm,draft=false]{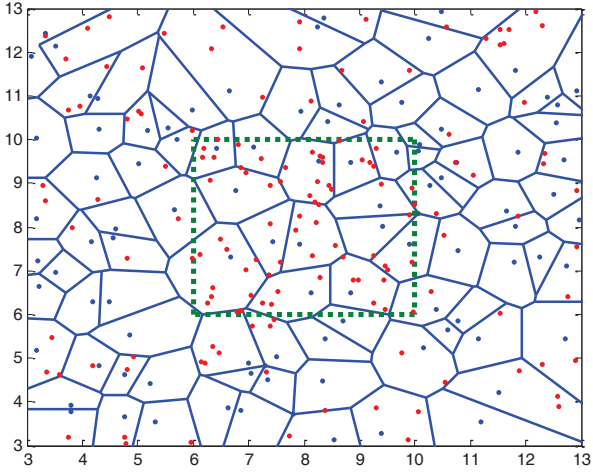}}
\caption{\small System model. The community is limited in rectangle plotted by the green line, blue nodes are macro cell BSs and blue lines are the coverage boundaries of macro cells, the red nodes are the small cell BSs.}
\end{figure}

\subsection{Individual Mobility Model}
In this paper the IMM is first introduced to evaluate the user mobility performance in 5G small cell networks. Differing with traditional user mobility models, the human tendency and clustering behaviors has been considered for IMM \cite{Song10}.In practical human society, the user visiting frequencies at different locations present obviously difference \cite{Halepovic05,Bayir09}, i.e., the user trends to visit some locations where he/she usually visits in the past. In this case, the user does not move with a fully random model in the real scenarios.

One mobility of user is called as one jump in IMM and the detail user mobility regulation is explained as follows \cite{Brockmann06}:

The user has two potential active modes before the next jump:
\begin{enumerate}
\item The user will visit a new location where has never been visited. The probability visiting a new location is expressed as
\[{\operatorname{P} _{new}}\left( n \right) = \rho S{\left( n \right)^{ - \gamma }},\tag{1}\]
where $0 < \rho  \leqslant 1$ and $\gamma  > 0$ are fixed parameters which are related with user mobility habits, $S\left( n \right)$ is the number of visited locations before the   jump,   is the jump number.
\item The user will returned an old location where has been visited in the past. The probability visiting an old location is expressed as
\[{\operatorname{P} _{ret}}\left( n \right) = 1 - \rho S{\left( n \right)^{ - \gamma }}.\tag{2}\]
\end{enumerate}
The user will pause at a location for a waiting time   after the user complete a jump. The probability of waiting time is expressed by $\operatorname{P} \left( {\Delta t} \right) \sim {\left| {\Delta t} \right|^{ - 1 - \beta }}$, where ${0 < \beta  \leqslant 1}$ is a fixed parameter which is measured by empirical data.

Based on the user mobility regulation in IMM, the user visits more locations, i.e., the value of $S\left( n \right)$ becomes larger, the probability of the user returning an old location becomes larger.

\section{User Mobility Performance}
Based on the IMM, the user mobility performance is analyzed for 5G small cell networks in this section. Considering the community configuration in Fig. 1, a Lemma is proposed as follows.

\textbf{Lemma1}: When the user mobility follows the IMM, the probability that the user jumps into the community of 5G small cell networks is equal to ${{{S_c}} \mathord{\left/
 {\vphantom {{{S_c}} {{S_t}}}} \right.
 \kern-\nulldelimiterspace} {{S_t}}}$, which is independent on the jump number $n$.

\textbf{Proof}: Considering that the user mobility follows the IMM, the result that the user jumps into the community in the $n - th$ jump can be composed of two cases, which are depicted as follows.

Case 1: The user explores a new location in the $n - th$ jump and the new location is located in the community, the corresponding probability is derived by

\[P_{new}^{in}\left( n \right) = {\operatorname{P} _{new}}\left( n \right)\frac{{{S_c}}}{{{S_t}}} = \rho S{\left( n \right)^{ - \gamma }}\frac{{{S_c}}}{{{S_t}}}.\tag{3}\]

Case 2: The user returns an old location in the $n - th$ jump and the old location is located in the community, the corresponding probability is derived by

\[P_{ret}^{in}\left( n \right) = {\operatorname{P} _{ret}}\left( n \right)\frac{{{N_c}{S_c}}}{{{S_t}}}\sum\limits_{i = 1}^{S\left( n \right)} {\frac{{{k_i}\left( n \right)}}{{n - 1}}},\tag{4}\]

where ${k_i}\left( n \right)$ is the visiting number at the $i - th$ location in the $n - th$ jump. Accounting for ${\sum\limits_{i = 1}^{S\left( n \right)} {{k_i}\left( n \right)}  = n - 1}$, (4) can be further derived by

\[P_{new}^{in}\left( n \right) = {\operatorname{P} _{ret}}\left( n \right)\frac{{{S_c}}}{{{S_t}}}\sum\limits_{i = 1}^{S\left( n \right)} {\frac{{{k_i}\left( n \right)}}{{n - 1}}}  = \left( {1 - \rho S{{\left( n \right)}^{ - \gamma }}} \right)\frac{{{S_c}}}{{{S_t}}}.\tag{5}\]

When (3) and (5) are summed, the probability that the user jumps into the community in the $n - th$ jump is ${\operatorname{P} _{c,in}} = {{{S_c}} \mathord{\left/
 {\vphantom {{{S_c}} {{S_t}}}} \right.
 \kern-\nulldelimiterspace} {{S_t}}}$. In the end, the proof of Lemma 1 is completed.

Let $t\left( n \right)$ is the total time spending for the number of $n$ jumps. ${t_{c,in}}\left( n \right)$ is the time spending in the community for the number of $n$ jumps. Let $\psi \left( n \right) = \frac{{{t_{c,in}}\left( n \right)}}{{t\left( n \right)}}$ is the ratio between the time spending in the community and the total time spending for the number of $n$ jumps. ${t_{c,m}}\left( n \right)$ is the moving time spending in the community for the number of   $n$ jumps. ${t_{c,p}}\left( n \right)$ is the pause time spending in the community for the number of $n$ jumps. Therefore, the time spending in the community for the number of $n$ jumps is extended as

\[{t_{c,in}}\left( n \right) = {t_{c,m}}\left( n \right) + {t_{c,p}}\left( n \right) = {t_{c,m}}\left( n \right) + {n_{c,in}}\Delta {t_c},\tag{6}\]

where ${n_{c,in}}$ is the jump number falling into the community in the jump number $n$, $\Delta {t_c}$ is the waiting time a user spent at one location in the community. Based on the result in \cite{Brockmann06}, the distribution of waiting time $\Delta {t_c}$ is governed by $\operatorname{P} \left( {\Delta {t_c}} \right) \sim {\left| {\Delta {t_c}} \right|^{ - 1 - {\beta _c}}}$, $0 < {\beta _c} \leqslant 1$. Let the user arrival and departure probabilities of the community are ${\pi _{c,in}}$ and ${\pi _{c,out}}$ respectively, which are expressed as

\[{\pi _{c,in}} = \mathop {\lim }\limits_{n \to \infty } {\mathbb{E}}\left[ {\psi \left( n \right)} \right] = \mathop {\lim }\limits_{n \to \infty } {\mathbb{E}}\left[ {\frac{{{t_{c,in}}\left( n \right)}}{{t\left( n \right)}}} \right],\tag{7}\]

\[{\pi _{c,out}} = 1 - {\pi _{c,in}} = 1 - \mathop {\lim }\limits_{n \to \infty } {\mathbb{E}}\left[ {\frac{{{t_{c,in}}\left( n \right)}}{{t\left( n \right)}}} \right],\tag{8}\]

where ${\mathbb{E}}\left[ \cdot \right]$  is the expectation operation.

Let ${n_{i,o}}$ is the moving number departing from the community, ${n_{o,i}}$  is the moving number arriving at the community, ${n_{i,i}}$  is the moving number inside the community, ${n_{o,o}}$  is the moving number outside the community. Therefore, total jump number is expressed by $n = {n_{o,i}} + {n_{i,i}} + {n_{i,o}} + {n_{o,o}}$ and the jump number falling into the community is expressed by ${n_{c,in}} = {n_{o,i}} + {n_{i,i}}$ . The time spending in the community for the number of ${t_{c,in}}\left( n \right)$  jumps   is derived by (9)
\begin{figure*}[!t]
\[\begin{gathered}
  {t_{c,in}}\left( n \right) = \sum\limits_{i = 1}^{{n_{i,i}}} {\left[ {\frac{{\left\| {L\left( I \right) - L\left( {k,k \in \left[ {0,I - 1} \right]} \right)} \right\|}}{{\overline v }}{\mathbf{1}}\left( \mathbb{A} \right) + \frac{{\left\| {L\left( I \right) - L\left( {k,k \in \left[ {0,I - 1} \right]} \right)} \right\|}}{{\overline v }}{\mathbf{1}}\left( {\mathbb{B}} \right)} \right]}  \hfill \\
  \quad \quad \quad  + \sum\limits_{i = 1}^{{n_{o,i}}} {\left[ {\frac{{\left\| {L\left( I \right) - L\left( {k,k \in \left[ {0,I - 1} \right]} \right)} \right\|}}{{\overline v }}{\mathbf{1}}\left( {\mathbb{A}} \right) + \frac{{\left\| {L\left( I \right) - L\left( {k,k \in \left[ {0,I - 1} \right]} \right)} \right\|}}{{\overline v }}{\mathbf{1}}\left( {\mathbb{B}} \right)} \right]}  \hfill \\
  \quad \quad \quad  + \sum\limits_{i = 1}^{{n_{i,o}}} {\left[ {\frac{{\left\| {L\left( I \right) - L\left( {k,k \in \left[ {0,I - 1} \right]} \right)} \right\|}}{{\overline v }}{\mathbf{1}}\left( {\mathbb{A}} \right) + \frac{{\left\| {L\left( I \right) - L\left( {k,k \in \left[ {0,I - 1} \right]} \right)} \right\|}}{{\overline v }}{\mathbf{1}}\left( {\mathbb{B}} \right)} \right]}  \hfill \\
  \quad \quad \quad  + {n_{c,in}}\Delta {t_c} \hfill \\
\end{gathered}. \tag{9}\]
\end{figure*}
where $L\left( I \right)$ is the location where the user moves at the $I - th$ jump, $\left\| \cdot \right\|$ is the distance function between two locations, $\overline v$  is the average user moving velocity, ${\mathbf{1}}\left( \cdot \right)$ is the indicator function, which equals to 1 when the event inside the bracket is occurred and 0 otherwise. $\mathbb{A}$ is the event that the user explores a new location at the $I - th$ jump and $\mathbb{B}$ is the event that the user returns an old location at the $I - th$ jump. Similarly, the total time spending for the number of $n$  jumps   $t\left( n \right)$ is derived by (10)
\begin{figure*}[!t]
\[\begin{gathered}
  t\left( n \right) = \sum\limits_{i = 1}^{{n_{i,i}}} {\left[ {\frac{{\left\| {L\left( I \right) - L\left( {k,k \in \left[ {0,I - 1} \right]} \right)} \right\|}}{{\overline v }}{\mathbf{1}}\left( {\mathbb{A}} \right) + \frac{{\left\| {L\left( I \right) - L\left( {k,k \in \left[ {0,I - 1} \right]} \right)} \right\|}}{{\overline v }}{\mathbf{1}}\left( {\mathbb{B}} \right)} \right]}  \hfill \\
  \quad \quad \; + \sum\limits_{i = 1}^{{n_{o,i}}} {\left[ {\frac{{\left\| {L\left( I \right) - L\left( {k,k \in \left[ {0,I - 1} \right]} \right)} \right\|}}{{\overline v }}{\mathbf{1}}\left( {\mathbb{A}} \right) + \frac{{\left\| {L\left( I \right) - L\left( {k,k \in \left[ {0,I - 1} \right]} \right)} \right\|}}{{\overline v }}{\mathbf{1}}\left( {\mathbb{B}} \right)} \right]}  \hfill \\
  \quad \quad \; + \sum\limits_{i = 1}^{{n_{i,o}}} {\left[ {\frac{{\left\| {L\left( I \right) - L\left( {k,k \in \left[ {0,I - 1} \right]} \right)} \right\|}}{{\overline v }}{\mathbf{1}}\left( {\mathbb{A}} \right) + \frac{{\left\| {L\left( I \right) - L\left( {k,k \in \left[ {0,I - 1} \right]} \right)} \right\|}}{{\overline v }}{\mathbf{1}}\left( {\mathbb{B}} \right)} \right]}  \hfill \\
  \quad \quad \; + \sum\limits_{i = 1}^{{n_{o,o}}} {\left[ {\frac{{\left\| {L\left( I \right) - L\left( {k,k \in \left[ {0,I - 1} \right]} \right)} \right\|}}{{\overline v }}{\mathbf{1}}\left( {\mathbb{A}} \right) + \frac{{\left\| {L\left( I \right) - L\left( {k,k \in \left[ {0,I - 1} \right]} \right)} \right\|}}{{\overline v }}{\mathbf{1}}\left( {\mathbb{B}} \right)} \right]}  \hfill \\
  \quad \;\,\,\,\;\, + n\Delta t \hfill \\
\end{gathered}. \tag{10}\]
\end{figure*}
Based on the result in Appendix A, ${\pi _{c,in}}$  is derived by (11)
\begin{figure*}[!t]
{\small{
\[\begin{gathered}
\begin{aligned}
  {\pi _{c,in}} &= \mathop {\lim }\limits_{n \to \infty } {\mathbb{E}}\left[ {\frac{{\left( {{n_{i,i}} + {n_{o,i}} + {n_{i,o}}} \right)\left\| {L\left( I \right) - L\left( {k,k \in \left[ {0,I - 1} \right]} \right)} \right\| + \left( {{n_{i,i}} + {n_{o,i}}} \right)\Delta {t_c}}}{{\overline v t\left( n \right)}}} \right] \hfill \\
  & = \mathop {\lim }\limits_{n \to \infty } \frac{{n\frac{1}{{\overline v }}\left[ {{{\left( {\frac{{{S_c}}}{{{S_t}}}} \right)}^2}{\mathbb{E}}\left[ {{d_{i,i}}} \right] + \left( {1 - \frac{{{S_c}}}{{{S_t}}}} \right)\frac{{{S_c}}}{{{S_t}}}{\mathbb{E}}\left[ {{d_{o,i}}} \right]} \right] + \frac{{{S_c}}}{{{S_t}}}n{\mathbb{E}}\left[ {\Delta {t_c}} \right]}}{{n\frac{1}{{\overline v }}\left[ {{{\left( {\frac{{{S_c}}}{{{S_t}}}} \right)}^2}{\mathbb{E}}\left[ {{d_{i,i}}} \right] + 2\left( {1 - \frac{{{S_c}}}{{{S_t}}}} \right)\frac{{{S_c}}}{{{S_t}}}{\mathbb{E}}\left[ {{d_{o,i}}} \right] + {{\left( {1 - \frac{{{S_c}}}{{{S_t}}}} \right)}^2}{\mathbb{E}}\left[ {{d_{o,o}}} \right]} \right] + n\frac{{{S_c}}}{{{S_t}}}{\mathbb{E}}\left[ {\Delta {t_c}} \right] + n\left( {1 - \frac{{{S_c}}}{{{S_t}}}} \right){\mathbb{E}}\left[ {\Delta {t_s}} \right]}} \hfill \\
  & = \frac{{\frac{1}{{\overline v }}\left[ {{{\left( {\frac{{{S_c}}}{{{S_t}}}} \right)}^2}{\mathbb{E}}\left[ {{d_{i,i}}} \right] + \left( {1 - \frac{{{S_c}}}{{{S_t}}}} \right)\frac{{{S_c}}}{{{S_t}}}{\mathbb{E}}\left[ {{d_{o,i}}} \right]} \right] + \frac{{{S_c}}}{{{S_t}}}{\mathbb{E}}\left[ {\Delta {t_c}} \right]}}{{\frac{1}{{\overline v }}\left[ {{{\left( {\frac{{{S_c}}}{{{S_t}}}} \right)}^2}{\mathbb{E}}\left[ {{d_{i,i}}} \right] + 2\left( {1 - \frac{{{S_c}}}{{{S_t}}}} \right)\frac{{{S_c}}}{{{S_t}}}{\mathbb{E}}\left[ {{d_{o,i}}} \right] + {{\left( {1 - \frac{{{S_c}}}{{{S_t}}}} \right)}^2}{\mathbb{E}}\left[ {{d_{o,o}}} \right]} \right] + \frac{{{S_c}}}{{{S_t}}}{\mathbb{E}}\left[ {\Delta {t_c}} \right] + \left( {1 - \frac{{{S_c}}}{{{S_t}}}} \right){\mathbb{E}}\left[ {\Delta {t_s}} \right]}} \hfill \\
  \end{aligned}
\end{gathered}. \normalsize{\tag{11}}\]
}}
\end{figure*}
where ${d_{i,i}}$  is the distance between two points inside the community, ${d_{o,i}}$  is the distance between one point located inside the community and another point located outside the community, ${d_{o,o}}$  is the distance between two points outside the community, $\Delta {t_s}$ is the waiting time a user spent at one location outside the community, the distribution of $\Delta {t_s}$ is governed by $\operatorname{P} \left( {\Delta {t_s}} \right) \sim {\left| {\Delta {t_s}} \right|^{ - 1 - {\beta _s}}}$, $0 < {\beta _s} \leqslant 1$. Substitute (11) into (8), the user departure probability  ${\pi _{c,out}}$ can be obtained.

Based on the derivation method in (11), the user pause probability  , i.e., the probability that a user pauses at a location, is derived by (12)
\begin{figure*}[!t]
{\small{
\[\begin{gathered}
\begin{aligned}
  {\pi _{pause}} &= \mathop {\lim }\limits_{n \to \infty } \mathbb{E}\left( {\frac{{{t_p}\left( n \right)}}{{t\left( n \right)}}} \right) \hfill \\
  & = \frac{{\frac{{{S_c}}}{{{S_t}}}{\mathbb{E}}\left[ {\Delta {t_c}} \right] + \left( {1 - \frac{{{S_c}}}{{{S_t}}}} \right){\mathbb{E}}\left[ {\Delta {t_s}} \right]}}{{\frac{1}{{\overline v }}\left[ {{{\left( {\frac{{{S_c}}}{{{S_t}}}} \right)}^2}{\mathbb{E}}\left[ {{d_{i,i}}} \right] + 2\left( {1 - \frac{{{S_c}}}{{{S_t}}}} \right)\frac{{{S_c}}}{{{S_t}}}{\mathbb{E}}\left[ {{d_{o,i}}} \right] + {{\left( {1 - \frac{{{S_c}}}{{{S_t}}}} \right)}^2}{\mathbb{E}}\left[ {{d_{o,o}}} \right]} \right] + \frac{{{S_c}}}{{{S_t}}}{\mathbb{E}}\left[ {\Delta {t_c}} \right] + \left( {1 - \frac{{{S_c}}}{{{S_t}}}} \right){\mathbb{E}}\left[ {\Delta {t_s}} \right]}}\quad  \hfill \\
\end{aligned}
\end{gathered},\normalsize{\tag{12}}\]
}}
\end{figure*}
where ${t_p}\left( n \right)$ is the total pausing time for the number of $n$  jumps. $\mathbb{E}\left[ {{d_{i,i}}} \right]$, $\mathbb{E}\left[ {{d_{o,i}}} \right]$ and $\mathbb{E}\left[ {{d_{o,o}}} \right]$ in (11) and (12) are derived as follows. Considering the derivation method is same for $\mathbb{E}\left[ {{d_{i,i}}} \right]$ , $\mathbb{E}\left[ {{d_{o,i}}} \right]$ and $\mathbb{E}\left[ {{d_{o,o}}} \right]$ , we only depict the detail derivation process for $\mathbb{E}\left[ {{d_{i,i}}} \right]$  and then directly list the final derivation result for $\mathbb{E}\left[ {{d_{o,i}}} \right]$  and $\mathbb{E}\left[ {{d_{o,o}}} \right]$.

Considering that the community is a rectangle, the length and width of the community are denoted as ${l_c}$  and ${w_c}$ , respectively. Without loss of generality, the center point of rectangle $\mathbb{R}_c^2$ is configured as the original point of Euclid coordinate $\left( {0,0} \right)$. ${u_K}$ and ${u_M}$ are two independent random points in the community and are governed by two independent uniform distributions. Locations of ${u_K}$ and ${u_M}$ are denoted as ${u_K}\left( {{x_K},{y_K}} \right)$ and ${u_M}\left( {{x_M},{y_M}} \right)$, respectively. Moreover, ${x_1} \leqslant {x_K} \leqslant {x_2}$ and ${x_1} \leqslant {x_M} \leqslant {x_2}$, ${l_c} = \left| {{x_2} - {x_1}} \right|$. ${y_1} \leqslant {y_K} \leqslant {y_2}$, ${y_1} \leqslant {y_M} \leqslant {y_2}$ and ${w_c} = \left| {{y_2} - {y_1}} \right|$, where $x_1$, $y_1$, $x_2$, $y_2$ are the coordinates of the boundaries of community area $\mathbb{R}_c^2$. Therefore, the distance between two points inside the community is denoted as ${d_{i,i}} = \sqrt {{{\left( {{x_K} - {x_M}} \right)}^2} + {{\left( {{y_K} - {y_M}} \right)}^2}} $. The probability density function (PDF) of  ${d_{i,i}}$ is expressed as

\[\begin{gathered}
\begin{aligned}
  {f_{{d_{i,i}}}}\left[ x \right]& = \frac{{dP\left[ {\sqrt {{{\left( {{x_K} - {x_M}} \right)}^2} + {{\left( {{y_K} - {y_M}} \right)}^2}}  < x} \right]}}{{dx}} \hfill \\
  & = \frac{{dP\left[ {{{\left( {{x_K} - {x_M}} \right)}^2} + {{\left( {{y_K} - {y_M}} \right)}^2} < {x^2}} \right]}}{{dx}} \hfill \\
  & = \frac{{d\int_0^{{x^2}} {{f_{{{\left( {{x_K} - {x_M}} \right)}^2} + {{\left( {{y_K} - {y_M}} \right)}^2}}}\left[ X \right]dX} }}{{dx}} \hfill \\
  &= 2x{f_{{{\left( {{x_K} - {x_M}} \right)}^2} + {{\left( {{y_K} - {y_M}} \right)}^2}}}\left[ {{x^2}} \right] \hfill \\
\end{aligned}
\end{gathered}. \tag{13}\]

Let  ${\Phi _X}\left[ t \right]$ is the characteristic function of the random variable $X$ , the term of {\small{${f_{{{\left( {{x_K} - {x_M}} \right)}^2} + {{\left( {{y_K} - {y_M}} \right)}^2}}}\left[ {{x^2}} \right]$}}  in (13) is further derived by (14a)
\begin{figure*}[!t]
{\normalsize{
\[{f_{{{\left( {{x_K} - {x_M}} \right)}^2} + {{\left( {{y_K} - {y_M}} \right)}^2}}}\left[ {{x^2}} \right] = \frac{1}{{2\pi }}\int_{ - \infty }^\infty  {{\Phi _{{{\left( {{x_K} - {x_M}} \right)}^2}}}\left[ t \right]{\Phi _{{{\left( {{y_K} - {y_M}} \right)}^2}}}\left[ t \right]} {e^{ - jt{x^2}}}dt,\tag{14a}\]}}
\end{figure*}
with (14b) and (14c).
\begin{figure*}[!t]
\[\begin{gathered}
\begin{aligned}
  {\Phi _{{{\left( {{x_K} - {x_M}} \right)}^2}}}\left[ t \right]& = \int_0^\infty  {\frac{{dP\left[ {{{\left( {{x_K} - {x_M}} \right)}^2} < x} \right]}}{{dx}}{e^{jtx}}} dx \hfill \\
  & = \int_0^\infty  {\left[ {\frac{{d\int_{ - \sqrt x }^{\sqrt x } {{f_{{x_K} - {x_M}}}\left( X \right)dX} }}{{dx}}} \right]{e^{jtx}}} dx \hfill \\
  & = \int_0^\infty  {\left[ {\frac{{d\left[ {\int_{ - \sqrt x }^{\sqrt x } {\frac{1}{{2\pi }}\left[ {\int_{ - \infty }^\infty  {\Phi _{{x_K}}^c\left( t \right)\Phi _{ - {x_M}}^c\left( t \right){e^{ - jtX}}dt} } \right]dX} } \right]}}{{dx}}} \right]{e^{jtx}}} dx \hfill \\
 & = \int_0^\infty  {\left[ {\frac{{d\left[ {\int_{ - \sqrt x }^{\sqrt x } {\frac{1}{{2\pi }}\left[ {\int_{ - \infty }^\infty  {\frac{{\left( {{e^{jt{x_2}}} - {e^{jt{x_1}}}} \right)\left( {{e^{ - jt{x_1}}} - {e^{ - jt{x_2}}}} \right)}}{{{t^2}{{\left( {{x_2} - {x_1}} \right)}^2}}}{e^{ - jtX}}dt} } \right]dX} } \right]}}{{dx}}} \right]{e^{jtx}}} dx\quad \hfill \\
\end{aligned}
\end{gathered}, \tag{14b}\]
\end{figure*}
\begin{figure*}[!t]
\[\begin{gathered}
\begin{aligned}
  {\Phi _{{{\left( {{y_K} - {y_M}} \right)}^2}}}\left[ t \right]& = \int_0^\infty  {\left[ {\frac{{d\left[ {\int_{ - \sqrt y }^{\sqrt y } {\frac{1}{{2\pi }}\left[ {\int_{ - \infty }^\infty  {\Phi _{{y_K}}^c\left( t \right)\Phi _{ - {y_M}}^c\left( t \right){e^{ - jtY}}dt} } \right]dY} } \right]}}{{dy}}} \right]{e^{jty}}} dy \hfill \\
  & = \int_0^\infty  {\left[ {\frac{{d\left[ {\int_{ - \sqrt y }^{\sqrt y } {\frac{1}{{2\pi }}\left[ {\int_{ - \infty }^\infty  {\frac{{\left( {{e^{jt{y_2}}} - {e^{jt{y_1}}}} \right)\left( {{e^{ - jt{y_1}}} - {e^{ - jt{y_2}}}} \right)}}{{{t^2}{{\left( {{y_2} - {y_1}} \right)}^2}}}{e^{ - jtY}}dt} } \right]dY} } \right]}}{{dy}}} \right]{e^{jty}}} dy\quad \hfill \\
\end{aligned}
\end{gathered}. \tag{14c}\]
\end{figure*}
Based on (13) and (14), the expectation of the distance  ${d_{i,i}}$ is given by (15).
\begin{figure*}[!t]
\[\begin{gathered}
\begin{aligned}
  \mathbb{E}\left[ {{d_{i,i}}} \right]& = \int_0^\infty  {x{f_{{d_{i,i}}}}\left[ x \right]} dx \hfill \\
  &= \int\limits_0^\infty  {\frac{{{x^2}}}{{4{\pi ^3}}}\int\limits_0^\infty  {\left[ {\int\limits_0^\infty  {\left[ {\frac{{d\left[ {\int_{ - \sqrt x }^{\sqrt x } {\left[ {\int_{ - \infty }^\infty  {\frac{{\left( {{e^{jt{x_2}}} - {e^{ - jt{x_1}}}} \right)\left( {{e^{jt{x_2}}} - {e^{jt{x_1}}}} \right)}}{{{t^2}{{\left( {{x_2} - {x_1}} \right)}^2}}}{e^{ - jtX}}dt} } \right]dX} } \right]}}{{dx}}} \right]{e^{jtx}}} dx} \right]} }  \hfill \\
  &\left[ {\int_0^\infty  {\left[ {\frac{{d\left[ {\int_{ - \sqrt y }^{\sqrt y } {\left[ {\int_{ - \infty }^\infty  {\frac{{\left( {{e^{jt{y_2}}} - {e^{ - jt{y_1}}}} \right)\left( {{e^{jt{y_2}}} - {e^{jt{y_1}}}} \right)}}{{{t^2}{{\left( {{y_2} - {y_1}} \right)}^2}}}{e^{ - jtY}}dt} } \right]dY} } \right]}}{{dy}}} \right]{e^{jty}}} dy} \right]{e^{ - jt{x^2}}}dtdx\quad \hfill \\
\end{aligned}
\end{gathered}. \tag{15}\]
\end{figure*}

Without loss of generality, the length and width of finite plane $\mathbb{R}_t^2$ are configured as ${l_t}$ and ${w_t}$, respectively. The center point of finite plane $\mathbb{R}_t^2$ is also configured at $\left( {0,0} \right)$. The location of a point ${u_H}$ is denoted as ${u_H}\left( {{x_H},{y_H}} \right)$, where ${x_3} \leqslant {x_H} \leqslant {x_4}$  and ${y_3} \leqslant {y_H} \leqslant {y_4}$ , ${l_t} = \left| {{x_4} - {x_3}} \right|$ and ${w_t} = \left| {{y_4} - {y_3}} \right|$ . To simplify the derivation, two formulations are configured as (16) and (17).
\begin{figure*}[!t]
\[{f_\Phi }\left( {a,b,c,d} \right) = \int\limits_0^\infty  {\left[ {\frac{{d\left[ {\int_{ - \sqrt x }^{\sqrt x } {\left[ {\int_{ - \infty }^\infty  {\frac{{\left( {{e^{jtb}} - {e^{jta}}} \right)\left( {{e^{ - jtc}} - {e^{ - jtd}}} \right)}}{{{t^2}\left( {b - a} \right)\left( {d - c} \right)}}{e^{ - jtX}}dt} } \right]dX} } \right]}}{{dx}}} \right]{e^{jtx}}} dx,\tag{16}\]
\end{figure*}

\begin{figure*}[!t]
\[\varsigma _{{x_{m,n}}}^{{y_{a,b}}} = \frac{{\left| {\left( {{x_n} - {x_m}} \right)\left( {{y_b} - {y_a}} \right)} \right|}}{{\left( {{y_4} - {y_3}} \right)\left( {{x_4} - {x_2}} \right) - \left( {{y_2} - {y_1}} \right)\left( {{x_2} - {x_1}} \right)}}.\tag{17}\]
\end{figure*}

Based on the derivation method for (15),  $\mathbb{E}\left[ {{d_{o,i}}} \right]$ and $\mathbb{E}\left[ {{d_{o,o}}} \right]$  are derived as (18) and (19).
\begin{figure*}[!t]
{\small{
\[\begin{gathered}
  \mathbb{E}\left[ {{d_{o,i}}} \right] = \int\limits_0^\infty  {\frac{{{x^2}}}{{4{\pi ^3}}}\int\limits_0^\infty  {\left[ {\varsigma _{{x_{2,4}}}^{{y_{3,4}}}{f_\Phi }\left( {{x_2},{x_4},{x_1},{x_2}} \right){f_\Phi }\left( {{y_3},{y_4},{y_1},{y_2}} \right)} \right. + \,} } \varsigma _{{x_{2,3}}}^{{y_{2,3}}}{f_\Phi }\left( {{x_2},{x_3},{x_1},{x_2}} \right){f_\Phi }\left( {{y_2},{y_3},{y_1},{y_2}} \right) \hfill \\
  \quad \quad \quad  + \,\varsigma _{{x_{1,3}}}^{{y_{1,4}}}{f_\Phi }\left( {{x_1},{x_3},{x_1},{x_2}} \right){f_\Phi }\left( {{y_1},{y_4},{y_1},{y_2}} \right) + \left. {\varsigma _{{x_{1,2}}}^{{y_{2,4}}}{f_\Phi }\left( {{x_1},{x_2},{x_1},{x_2}} \right){f_\Phi }\left( {{y_2},{y_4},{y_1},{y_2}} \right)} \right]{e^{ - jt{x^2}}}dtdx \hfill \\
\end{gathered}, \normalsize{\tag{18}}\]
}}
\end{figure*}

\begin{figure*}[!t]
{\small{
\[\begin{gathered}
  \mathbb{E}\left[ {{d_{o,o}}} \right] = \int\limits_0^\infty  {\frac{{{x^2}}}{{4{\pi ^3}}}\int\limits_0^\infty  {\left[ {\varsigma _{{x_{2,4}}}^{{y_{3,4}}}\varsigma _{{x_{2,4}}}^{{y_{3,4}}}{f_\Phi }\left( {{x_2},{x_4},{x_2},{x_4}} \right){f_\Phi }\left( {{y_3},{y_4},{y_3},{y_4}} \right)} \right. + \,} } \varsigma _{{x_{2,3}}}^{{y_{2,3}}}\varsigma _{{x_{2,3}}}^{{y_{2,3}}}{f_\Phi }\left( {{x_2},{x_3},{x_2},{x_3}} \right){f_\Phi }\left( {{y_2},{y_3},{y_2},{y_3}} \right) \hfill \\
  \quad \quad \quad  + \varsigma _{{x_{2,4}}}^{{y_{3,4}}}\varsigma _{{x_{1,3}}}^{{y_{1,4}}}{f_\Phi }\left( {{x_2},{x_4},{x_1},{x_3}} \right){f_\Phi }\left( {{y_3},{y_4},{y_1},{y_4}} \right) + \varsigma _{{x_{2,4}}}^{{y_{3,4}}}\varsigma _{{x_{2,3}}}^{{y_{2,3}}}{f_\Phi }\left( {{x_2},{x_4},{x_2},{x_3}} \right){f_\Phi }\left( {{y_3},{y_4},{y_2},{y_3}} \right) \hfill \\
  \quad \quad \quad  + \varsigma _{{x_{2,4}}}^{{y_{3,4}}}\varsigma _{{x_{1,2}}}^{{y_{2,4}}}{f_\Phi }\left( {{x_2},{x_4},{x_1},{x_2}} \right){f_\Phi }\left( {{y_3},{y_4},{y_2},{y_4}} \right) + \varsigma _{{x_{2,3}}}^{{y_{2,3}}}\varsigma _{{x_{1,3}}}^{{y_{1,4}}}{f_\Phi }\left( {{x_2},{x_3},{x_1},{x_3}} \right){f_\Phi }\left( {{y_2},{y_3},{y_1},{y_4}} \right) \hfill \\
  \quad \quad \quad  + \varsigma _{{x_{2,3}}}^{{y_{2,3}}}\varsigma _{{x_{1,2}}}^{{y_{2,4}}}{f_\Phi }\left( {{x_2},{x_3},{x_1},{x_2}} \right){f_\Phi }\left( {{y_2},{y_3},{y_2},{y_4}} \right) + \varsigma _{{x_{1,2}}}^{{y_{2,4}}}\varsigma _{{x_{1,3}}}^{{y_{1,4}}}{f_\Phi }\left( {{x_1},{x_2},{x_1},{x_3}} \right){f_\Phi }\left( {{y_2},{y_4},{y_1},{y_4}} \right) \hfill \\
  \quad \quad \quad  + \,\varsigma _{{x_{1,3}}}^{{y_{1,4}}}\varsigma _{{x_{1,3}}}^{{y_{1,4}}}{f_\Phi }\left( {{x_1},{x_3},{x_1},{x_2}} \right){f_\Phi }\left( {{y_1},{y_4},{y_1},{y_2}} \right) + \left. {\varsigma _{{x_{1,2}}}^{{y_{2,4}}}\varsigma _{{x_{1,2}}}^{{y_{2,4}}}{f_\Phi }\left( {{x_1},{x_2},{x_1},{x_2}} \right){f_\Phi }\left( {{y_2},{y_4},{y_1},{y_2}} \right)} \right]{e^{ - jt{x^2}}}dtdx\quad  \hfill \\
\end{gathered}. \normalsize{\tag{19}}\]
}}
\end{figure*}
Substitute (15), (18) and (19) into (11) and (12), the user arrival and pause probabilities, i.e., ${\pi _{c,in}}$  and ${\pi _{pause}}$  can be obtained in the end.

\section{Coverage Model}
Based on the group cell scheme, a static user $UE$  can associate with multiply small cell BSs if SINRs over wireless links are larger than or equal to a given threshold ${\gamma _0}$.  Assume that SINRs over wireless links are independent each other, the selection process of small cell BSs associated with a static user can be denoted as ${N_b}$  independent Bernoulli experiments, i.e.,

\[{N_b} = \left\{ \begin{gathered}
  {\lambda _{c,BS}}{S_c},\quad UE \in \mathbb{R}_c^2 \hfill \\
  {\lambda _{s,BS}}{S_s},\quad UE \in \mathbb{R}_s^2 \hfill \\
\end{gathered}.  \right.\tag{20}\]

In this case, the number of small cell BSs associating with a static user   inside and outside the community is denoted as ${n_{c,s}}$  and ${n_{s,s}}$ , respectively.  Moreover, ${n_{c,s}}$  and ${n_{s,s}}$ are independent each other and are governed by binomial distributions. Let coverage probabilities of a small cell BS inside and outside the community are  ${P_{c\_cover}}$ and ${P_{s\_cover}}$ , respectively. Probabilities of ${n_{c,s}}$  and ${n_{s,s}}$ inside and outside the community are expressed as (21) and (22).
\begin{figure*}[!t]
\[{\text{P}}\left( {{n_{c,s}}} \right) = \left( \begin{gathered}
  \left\lfloor {{\lambda _{c,BS}}{S_c}} \right\rfloor  \hfill \\
  \quad {n_{c,s}} \hfill \\
\end{gathered}  \right){\left( {{P_{c\_cover}}} \right)^{{n_{c,s}}}}{\left( {1 - {P_{c\_cover}}} \right)^{\left\lfloor {{\lambda _{c,BS}}{S_c}} \right\rfloor  - {n_{c,s}}}}\quad ,\tag{21}\]
\end{figure*}
\begin{figure*}[!t]
\[{\text{P}}\left( {{n_{s,s}}} \right) = \left( \begin{gathered}
  \left\lfloor {{\lambda _{s,BS}}{S_s}} \right\rfloor  \hfill \\
  \quad {n_{s,s}} \hfill \\
\end{gathered}  \right){\left( {{P_{s\_cover}}} \right)^{{n_{s,s}}}}{\left( {1 - {P_{s\_cover}}} \right)^{\left\lfloor {{\lambda _{s,BS}}{S_s}} \right\rfloor  - {n_{s,s}}}}\quad ,\tag{22}\]
\end{figure*}
where $\left\lfloor \cdot  \right\rfloor $  is the bottom integral function, $\tiny{\left( \begin{gathered}
  M \hfill \\
  N \hfill \\
\end{gathered}  \right)}$  is the binomial coefficients meaning the number of ways of picking  $N$ unordered outcomes from  $M$ possibilities. Considering the OFDM scheme is adopted at small cell BSs, the interference received at the static user is only generated from adjacent small cell BSs with the co-channel. The coverage probability of a small cell BS inside the community is expressed by (23),
\begin{figure*}[!t]
\[{P_{c\_cover}} = {\text{P}}\left[ {Recieved\;SINR > {\gamma _0}} \right] = \operatorname{P} \left[ {\frac{{{P_{rs}}}}{{{\sigma ^2} + \sum\limits_{k = 0}^{\left\lfloor {{\lambda _{c,BS}}{S_c} - {N_{cover}}} \right\rfloor } {{P_{ri}}} }}}>0 \right]\quad ,\tag{23}\]
\end{figure*}
where ${\gamma _0}$  is the SINR threshold over wireless links, ${\sigma ^2}$  is the Gaussian noise power, ${P_{rs}}$  and ${P_{ri}}$  are the desired signal power and the interference power at the static user, and ${N_{cover}}$ denotes the total number of available small cell BSs. In this paper the wireless channel is assumed to be the Rayleigh fading channel \cite{Park09}. Therefore, (23) is further derived by

\[{P_{c\_cover}} = \operatorname{P} \left[ {\frac{{{P_{ts}}{h_s}{d_s}^{ - \alpha }}}{{{\sigma ^2} + \sum\limits_{k = 0}^{\left\lfloor {{\lambda _{c,BS}}{S_c} - {N_{cover}}} \right\rfloor } {{P_{tin}}{h_{in}}{d_{in}}^{ - \alpha }} }} > {\gamma _0}} \right]\quad ,\tag{24}\]
where  ${P_{ts}}$ and ${P_{ti}}$  are the transmission power from the desired small cell BS and the interfering small cell BSs, respectively.  ${h_s}$  and ${h_in}$  are the small scale fading over the desired wireless channel and the interfering wireless channels. Moreover, ${h_s}$  and ${h_in}$    are independent each other and governed by an exponential distribution with the mean ${\lambda _h}$ . ${d_s}$  is the distance between the static user and the desired small cell BS.  ${d_{in}}$  is the distance between the static user and the interfering small cell BS. Considering that locations of the static user and the small cell BSs are governed by two independent uniform distributions,  ${d_s}$ and ${d_i}$  are assumed to be independent and identically distributed (i.i.d.) random variables. Assume that small cell BSs have the equal transmission power ${P_t}$, i.e.,  ${P_t} = {P_{ts}} = {P_{tin}}$ (24) can be rewritten by (25).
\begin{figure*}[!t]
\[{P_{c\_cover}} = \operatorname{P} \left[ {{P_t}{h_s}{d_s}^{ - \alpha } - {\gamma _0}{\sigma ^2} - {\gamma _0}\sum\limits_{k = 0}^{\left\lfloor {{\lambda _{c,BS}}{S_c} - {N_{cover}}} \right\rfloor } {{P_t}{h_{in}}{d_{in}}^{ - \alpha }}  > 0} \right]\quad .\tag{25}\]
\end{figure*}

Considering that three terms in (25), i.e., ${P_t}{h_s}{d_s}^{ - \alpha }$  , ${\gamma _0}{\sigma ^2}$  and ${\gamma _0}\sum\limits_{k = 0}^{\left\lfloor {{\lambda _{c,BS}}{S_c} - {N_{cover}}} \right\rfloor } {{P_t}{h_{in}}{d_{in}}^{ - \alpha }} $ are independent random variables, (25) can be calculated by the joint probability of three independent random variables.

Considering that ${h_s}$  and  ${h_{in}}$ are governed by an exponential distribution with the mean ${\lambda _h}$  , the PDF of ${h_s}$  and  ${h_{in}}$is expressed by

\[{f_h}\left( x \right) = {\lambda _h}{e^{ - {\lambda _h}x}}.\tag{26}\]

Furthermore, the PDF of the variable ${P_t}{h_s}$  is expressed by

\[\begin{gathered}
\begin{aligned}
  {f_{{P_t}{h_s}}}\left( x \right)& = \frac{{dP\left[ {{P_t}{h_s} < x} \right]}}{{dx}} = \frac{{dP\left[ {{h_s} < \frac{x}{{{P_t}}}} \right]}}{{dx}} \hfill \\
  &= \frac{{d\int_0^{\frac{x}{{{P_t}}}} {{\lambda _h}{e^{ - {\lambda _h}X}}dX} }}{{dx}} = \frac{1}{{{P_t}}}{f_h}\left( {\frac{x}{{{P_t}}}} \right) = \frac{{{\lambda _h}}}{{{P_t}}}{e^{ - \frac{{{\lambda _h}}}{{{P_t}}}x}} \hfill \\
\end{aligned}
\end{gathered}. \tag{27}\]

Based on the result of (13), the PDF of ${d_s}$  and ${d_in}$  is expressed by (28).
\begin{figure*}[!t]
\[{f_D}\left( x \right) = {f_{\sqrt {{{\left( {{x_K} - {x_M}} \right)}^2} + {{\left( {{y_K} - {y_M}} \right)}^2}} }}\left( x \right) = \left\{ \begin{gathered}
  {f_{{d_{i,i}}}}\left( x \right),\quad {x_K},{x_M},{y_K},{y_M} \in \mathbb{R}_c^2 \hfill \\
  {f_{{d_{o,o}}}}\left( x \right),\;\;\;{x_K},{x_M},{y_K},{y_M} \in \mathbb{R}_s^2 \hfill \\
\end{gathered}  \right.\quad .\tag{28}\]
\end{figure*}
Furthermore, the PDF of the variable ${d_s}^{ - \alpha }$  is expressed by

\[{f_{{d_s}^{ - \alpha }}}\left( x \right) = \frac{{d\int_0^{{x^\alpha }} {{f_D}\left( X \right)dX} }}{{dx}} = \alpha {x^{\alpha  - 1}}{f_D}\left( {{x^\alpha }} \right).\tag{29}\]

Based on (27) and (29), the characteristic function of ${P_t}{h_s}{d_s}^{ - \alpha }$  is derived by

\[\begin{gathered}
\begin{aligned}
  {\Phi _{{P_t}{h_s}{d_s}^{ - \alpha }}}\left[ t \right] &= \int_0^\infty  {\frac{1}{{{P_t}}}{f_h}\left( {\frac{x}{{{P_t}}}} \right) \cdot {f_{d_s^{ - \alpha }}}\left( x \right) \cdot {e^{jtx}}dx}  \hfill \\
  &= \int_0^\infty  {\frac{{{\lambda _h}}}{{{P_t}}}{e^{ - \frac{{{\lambda _h}}}{{{P_t}}}x}} \cdot \alpha {x^{\alpha  - 1}}{f_D}\left( {{x^\alpha }} \right) \cdot {e^{jtx}}dx}  \hfill \\
  \end{aligned}
\end{gathered}. \tag{30}\]

Based on the same derivation method used for (30), the PDF of $ - {\gamma _0}{P_t}{h_{in}}{d_{in}}^{ - \alpha }$ is derived by (31).
\begin{figure*}[!t]
\[\begin{gathered}
\begin{aligned}
  {f_{ - {\gamma _0}{P_t}{h_{in}}{d_{in}}^{ - \alpha }}}\left[ x \right]& = \frac{{dP\left[ { - {\gamma _0}{P_t}{h_{in}}{d_{in}}^{ - \alpha } < x} \right]}}{{dx}} = \frac{{dP\left[ {{P_t}{h_{in}}{d_{in}}^{ - \alpha } >  - \frac{x}{{{\gamma _0}}}} \right]}}{{dx}} \hfill \\
  &= \frac{{d\left[ {1 - P\left[ {{P_t}{h_{in}}{d_{in}}^{ - \alpha } <  - \frac{x}{{{\gamma _0}}}} \right]} \right]}}{{dx}} =  - \frac{{dP\left[ {{P_t}{h_{in}}{d_{in}}^{ - \alpha } <  - \frac{x}{{{\gamma _0}}}} \right]}}{{dx}} \hfill \\
  &=  - \frac{{d\int_0^{ - \frac{x}{{{\gamma _0}}}} {\frac{{{\lambda _h}}}{{{P_t}}}{e^{ - \frac{{{\lambda _h}}}{P}X}} \cdot \alpha {X^{\alpha  - 1}}{f_{{d_{i,i}}}}\left[ {{X^\alpha }} \right]} dX}}{{dx}} \hfill \\
  &= \frac{1}{{{\gamma _0}}}\frac{{{\lambda _h}}}{{{P_t}}}{e^{\frac{{{\lambda _h}x}}{{{P_t}{\gamma _0}}}}} \cdot \alpha {\left( { - \frac{x}{{{\gamma _0}}}} \right)^{\alpha  - 1}}{f_{{d_{i,i}}}}\left[ {{{\left( { - \frac{x}{{{\gamma _0}}}} \right)}^\alpha }} \right]\quad ,\quad x \in \left( { - \infty ,0} \right] \hfill \\
  \end{aligned}
\end{gathered}. \tag{31}\]
\end{figure*}
Furthermore, the characteristic function of $ - {\gamma _0}{P_t}{h_{in}}{d_{in}}^{ - \alpha }$  is derived by (32).
\begin{figure*}[!t]
\[{\Phi _{ - {\gamma _0}{P_t}{h_{in}}{d_{in}}^{ - \alpha }}}\left[ t \right] = \int_{ - \infty }^0 {\frac{1}{{{\gamma _0}}}\frac{{{\lambda _h}}}{{{P_t}}}{e^{\frac{{{\lambda _h}x}}{{{P_t}{\gamma _0}}}}} \cdot \alpha {{\left( { - \frac{x}{{{\gamma _0}}}} \right)}^{\alpha  - 1}}{f_{{d_{i,i}}}}\left[ {{{\left( { - \frac{x}{{{\gamma _0}}}} \right)}^\alpha }} \right] \cdot {e^{jtx}}dx} \quad .\tag{32}\]
\end{figure*}
Assume that the noise over wireless channels is governed by the standard normal distribution. The PDF of $ - {\gamma _0}{\sigma ^2}$  is derived by (33).
\begin{figure*}[!t]
\[\begin{gathered}
\begin{aligned}
  {f_{ - {\sigma ^2}{\gamma _0}}}\left[ x \right]& = \frac{{dP\left[ { - {\sigma ^2}{\gamma _0} < x} \right]}}{{dx}} = \frac{{dP\left[ {{\sigma ^2} >  - \frac{x}{{{\gamma _0}}}} \right]}}{{dx}} = \frac{{d\left[ {1 - P\left[ {{\sigma ^2} <  - \frac{x}{{{\gamma _0}}}} \right]} \right]}}{{dx}} \hfill \\
  &=  - \frac{{dP\left[ {{\sigma ^2} <  - \frac{x}{{{\gamma _0}}}} \right]}}{{dx}} =  - \frac{{dP\left[ { - \sqrt { - \frac{x}{{{\gamma _0}}}}  < \sigma  < \sqrt { - \frac{x}{{{\gamma _0}}}} } \right]}}{{dx}} \hfill \\
  &=  - \frac{{d\int_{ - \sqrt { - \frac{x}{{{\gamma _0}}}} }^{\sqrt { - \frac{x}{{{\gamma _0}}}} } {\frac{1}{{\sqrt {2\pi } \sigma }}} {e^{ - \frac{{{X^2}}}{{2{\sigma ^2}}}}}dX}}{{dx}} = \frac{1}{{{\gamma _0}\sigma \sqrt { - \frac{{2\pi x}}{{{\gamma _0}}}} }}{e^{\frac{x}{{2{\gamma _0}{\sigma ^2}}}}}\quad ,\quad x \in \left( { - \infty ,0} \right] \hfill \\
\end{aligned}
\end{gathered}. \tag{33}\]
\end{figure*}
Furthermore, the characteristic function of $ - {\gamma _0}{\sigma ^2}$  is derived by (34).
\begin{figure*}[!t]
\[{\Phi _{ - {\sigma ^2}{\gamma _0}}}\left[ t \right] = \int_{ - \infty }^0 {\frac{1}{{{\gamma _0}\sigma \sqrt { - \frac{{2\pi x}}{{{\gamma _0}}}} }}{e^{\frac{x}{{2{\gamma _0}{\sigma ^2}}}}} \cdot {e^{jtx}}dx}  = \int_{ - \infty }^0 {\frac{1}{{{\gamma _0}\sigma \sqrt { - \frac{{2\pi x}}{{{\gamma _0}}}} }}{e^{\frac{{1 + 2jt{\gamma _0}{\sigma ^2}}}{{2{\gamma _0}{\sigma ^2}}}x}}dx} \quad .\tag{34}\]
\end{figure*}
Based on (30), (32) and (34), the characteristic function of{\footnotesize{ $\xi  = {P_t}{h_s}{D_s}^{ - \alpha } - {\sigma ^2}{\gamma _0} - {\gamma _0}\sum\limits_{k = 0}^{\left\lfloor {{\lambda _{c,BS}}{S_c} - {N_{cover}}} \right\rfloor } {{P_t}{h_{in}}{d_{in}}^{ - \alpha }} $}} is derived by (35).
\begin{figure*}[!t]
\[\begin{gathered}
\begin{aligned}
  {\Phi _\xi }\left[ t \right] &= {\Phi _{{P_t}{h_s}{d_s}^{ - \alpha }}}\left[ t \right] \cdot {\Phi _{ - {\sigma ^2}{\gamma _0}}}\left[ t \right] \cdot {\left[ {{\Phi _{ - {\gamma _0}{P_t}{h_i}{d_i}^{ - \alpha }}}\left[ t \right]} \right]^{\left\lfloor {{\lambda _{c,BS}}{S_c} - {N_{cover}}} \right\rfloor }} \hfill \\
  &= \left[ {\int_0^\infty  {\frac{{{\lambda _h}}}{{{P_t}}}{e^{ - \frac{{{\lambda _h}}}{{{P_t}}}x}} \cdot \alpha {x^{\alpha  - 1}}{f_{{d_{i,i}}}}\left[ {{x^\alpha }} \right] \cdot {e^{jtx}}dx} } \right] \hfill \\
   &\cdot \left[ {\int_{ - \infty }^0 {\frac{1}{{{\gamma _0}\sigma \sqrt { - \frac{{2\pi x}}{{{\gamma _0}}}} }}{e^{\frac{{1 + 2jt{\gamma _0}{\sigma ^2}}}{{2{\gamma _0}{\sigma ^2}}}x}}dx} } \right] \hfill \\
   &\cdot {\left[ {\int_{ - \infty }^0 {\frac{1}{{{\gamma _0}}}\frac{{{\lambda _h}}}{{{P_t}}}{e^{\frac{{{\lambda _h}x}}{{{P_t}{\gamma _0}}}}} \cdot \alpha {{\left( { - \frac{x}{{{\gamma _0}}}} \right)}^{\alpha  - 1}}{f_{{d_{i,i}}}}\left[ {{{\left( { - \frac{x}{{{\gamma _0}}}} \right)}^\alpha }} \right] \cdot {e^{jtx}}dx} } \right]^{\left\lfloor {{\lambda _{c,BS}}{S_c} - {N_{cover}}} \right\rfloor }} \hfill \\
  \end{aligned}
\end{gathered}. \tag{35}\]
\end{figure*}
Base on the Fourier inversion transform, the PDF of $\xi $  is derived by

\[{f_\xi }\left[ x \right] = \frac{1}{{2\pi }}\int_{ - \infty }^\infty  {{\Phi _\xi }\left[ t \right]}  \cdot {e^{ - jtx}}dt\quad ,\tag{36}\]

Based on (25) and (36), the coverage probability of a small cell BS for a static user inside the community is derived by

\[{P_{c\_cover}} = 1 - \frac{1}{{2\pi }}\int_{ - \infty }^0 {\left( {\int_{ - \infty }^\infty  {{\Phi _\xi }\left[ t \right]}  \cdot {e^{ - jtx}}dt} \right)\,} dx\quad ,\tag{37}\]

Based on the derivation method used for (37), the coverage probability of a small cell BS for a static user outside the community is derived by (38a)
\begin{figure*}[!t]
\[{P_{s\_cover}} = 1 - \frac{1}{{2\pi }}\int_{ - \infty }^0 {\left( {\int_{ - \infty }^\infty  {\Phi' _\xi\left[ t \right]}  \cdot {e^{ - jtx}}dt} \right)\,} dx\quad , \tag{38a}\]
\end{figure*}
with (38b)
\begin{figure*}[!t]
\[\begin{gathered}
  \Phi' _\xi \left[ t \right] = \left[ {\int_0^\infty  {\frac{{{\lambda _h}}}{{{P_t}}}{e^{ - \frac{{{\lambda _h}}}{{{P_t}}}x}} \cdot \alpha {x^{\alpha  - 1}}{f_{{d_{o,o}}}}\left( {{x^\alpha }} \right) \cdot {e^{jtx}}dx} } \right] \cdot \left[ {\int_{ - \infty }^0 {\frac{1}{{{\gamma _0}\sigma \sqrt { - \frac{{2\pi x}}{{{\gamma _0}}}} }}{e^{\frac{{1 + 2jt{\gamma _0}{\sigma ^2}}}{{2{\gamma _0}{\sigma ^2}}}x}}dx} } \right] \hfill \\
  \quad \quad \quad \; \cdot {\left[ {\int_{ - \infty }^0 {\frac{1}{{{\gamma _0}}}\frac{{{\lambda _h}}}{{{P_t}}}{e^{\frac{{{\lambda _h}x}}{{{P_t}{\gamma _0}}}}} \cdot \alpha {{\left( { - \frac{x}{{{\gamma _0}}}} \right)}^{\alpha  - 1}}{f_{{d_{o,o}}}}\left( {{{\left( { - \frac{x}{{{\gamma _0}}}} \right)}^\alpha }} \right) \cdot {e^{jtx}}dx} } \right]^{\left\lfloor {{\lambda _{s,BS}}{S_c} - {N_{cover}}} \right\rfloor }}\quad  \hfill \\
\end{gathered}. \tag{38b}\]
\end{figure*}

Based on (11), (12), (37) and (38), the number of available small cell BSs for a static user is derived by (39a)

\[{N_{cover}} = \left( {{\pi _{c,in}}\mathbb{E}\left[ {{n_{c,s}}} \right] + \left( {1 - {\pi _{c,in}}} \right)\mathbb{E}\left[ {{n_{s.s}}} \right]} \right){\pi _{pause}}\quad ,\tag{39a}\]
with
\[\mathbb{E}\left[ {{n_{c,s}}} \right] = {\lambda _{c,BS}}{S_c}{P_{c\_cover}}\quad,\tag{39b}\]
\[\mathbb{E}\left[ {{n_{s.s}}} \right] = {\lambda _{s,BS}}{S_s}{P_{s\_cover}}\quad.\tag{39c}\]

Considering that ${P_{c\_cover}}$ and ${P_{s\_cover}}$ both are functions of ${N_{cover}}$, The computation of ${N_{cover}}$ in (39) has to be performed by iteration calculations.

Differing with the coverage probability for the static user, the coverage probability of the moving user is defined as: when the user moves with the average velocity  $\overline v $ for a period $\Delta {t_m}$ , the moving user is covered by macro cell BSs only if the moving user SINR is always larger than or equal to a given threshold ${\gamma _0}$  in the period $\Delta {t_m}$ . Assume that the distance between the initial location of moving user and the associated macro cell BS is ${r_p} = \left| {\overrightarrow {{r_p}} } \right|$ , where $\overrightarrow {{r_p}} $  is the distance vector from the moving user to the associated macro cell BS.  The user moving distance is denoted as $\Delta {r_m} = \overline v  \cdot \Delta {t_m}$ . The included angle between the user moving direction and the distance vector $\overrightarrow {{r_p}} $  is dented as ${\theta _m}$ .  Hence, the distance between the location of moving user and the associated macro cell BS is  ${r_m} = \sqrt {r_p^2 + \Delta r_m^2 - 2{r_p}\Delta {r_m}\cos \left( {{\theta _m}} \right)} $  after the user moves a distance $\Delta {r_m}$ .Let the event   $\mathbb{C}$ is that the moving user is associated with the macro cell BS $MBS$  and the event $\mathbb{D}$  is that the distance between the location of moving user and the associated macro cell BS  $MBS$ is ${r_m}$  after the user moves a distance $\Delta {r_m}$ . Therefore, the coverage probability of a macro cell BS for the moving user is expressed by
\[{P_{m\_cover}} = {\text{P}}\left( {\mathbb{C}\left| \mathbb{D} \right.} \right)\quad .\tag{40}\]

Based on the result in \cite{Jeffrey11}, (40) can be extended as (41)
\begin{figure*}[!t]
\[{P_{m\_cover}} = \int\limits_{x > 0} {{f_{{r_m}}}\left( x \right)} \int\limits_{ - \infty }^\infty  {\frac{{{e^{ - 2\pi {\sigma ^2}sj}}{\mathcal{L}_{{I_r}}}\left( {2\pi sj} \right)\left[ {{\mathcal{L}_h}\left( { - 2\pi {{\left( {{\gamma _0}x} \right)}^{ - 1}}sj} \right) - 1} \right]}}{{2\pi sj}}} dsdx\quad ,\tag{41a}\]
\[{\mathcal{L}_{{I_r}}}\left( {2\pi sj} \right) = {e^{ - 2\pi x\int\limits_x^\infty  {\left( {1 - \frac{{{\lambda _h}}}{{{\lambda _h} + 2\pi s{v^{ - \alpha }}j}}} \right)vdv} }}\quad ,\tag{41b}\]
\[{\mathcal{L}_h}\left( { - 2\pi {{\left( {{\gamma _0}x} \right)}^{ - 1}}sj} \right) = \frac{{{\lambda _h}}}{{{\lambda _h} - 2\pi {{\left( {{\gamma _0}x} \right)}^{ - 1}}sj}}\quad ,\tag{41c}\]
\end{figure*}
where ${\mathcal{L}_{{I_r}}}\left( \cdot \right)$  is the Laplace transform of the interference at the moving user,   ${\mathcal{L}_{{h}}}\left( \cdot \right)$is the Laplace transform of the desired signal at the moving user.

Considering that the location of macro cell BS is governed by a Poisson point process distribution and the initial location of moving user is governed by a uniform distribution, the PDF of the distance  ${r_p}$ is expressed as \cite{Haenggi05}
\[{f_{{r_p}}}\left( x \right) = 2\pi {\lambda _{m,BS}}x{e^{ - \pi {\lambda _{m,BS}}{x^2}}}\quad .\tag{42}\]
Assume that ${\theta _m}$  is uniformly distributed in $\left[ {0,\pi } \right]$  and  $\Delta {r_m}$ is fixed, the PDF of  ${r_m}$ is derived by (43a)
\begin{figure*}[!t]
\[\begin{gathered}
\begin{aligned}
  {f_{{r_m}}}\left( x \right) &= \frac{{d{\text{P}}\left( {\sqrt {r_p^2 + \Delta r_m^2 - 2{r_p}\Delta {r_m}\cos \left( {{\theta _m}} \right)}  < x} \right)}}{{dx}} \hfill \\
  &= \frac{{d\left( {\frac{1}{\pi }\int\limits_0^\pi  {d{\theta _m}} \int\limits_0^{\Delta {r_m}\cos {\theta _m} + \sqrt {{x^2} + {{\left( {\Delta {r_m}\cos \left( {{\theta _m}} \right)} \right)}^2} - \Delta {r_m}^2} } {{f_{{r_p}}}\left( y \right)dy} } \right)}}{{dx}} \hfill \\
   &= \frac{{d\left( {\frac{1}{\pi }\int\limits_0^\pi  {d{\theta _m}} \int\limits_0^{\Delta {r_m}\cos {\theta _m} + \sqrt {{x^2} + {{\left( {\Delta {r_m}\cos \left( {{\theta _m}} \right)} \right)}^2} - \Delta {r_m}^2} } {2\pi {\lambda _m}y{e^{ - \pi {\lambda _m}{y^2}}}dy} } \right)}}{{dx}} \hfill \\
  &= \frac{1}{\pi }\int\limits_0^\pi  {\frac{{2\pi x{\lambda _m}{\omega _m}\left( {x,{\theta _m}} \right){e^{ - \pi {\lambda _m}{{\left( {{\omega _m}\left( {x,{\theta _m}} \right)} \right)}^2}}}}}{{{\omega _m}\left( {x,{\theta _m}} \right) - \Delta {r_m}\cos {\theta _m}}}d{\theta _m}}  \hfill \\
  \end{aligned}
\end{gathered}, \tag{43a}\]
\end{figure*}
with (43b)
\begin{figure*}[!t]
\[{\omega _m}\left( {x,{\theta _m}} \right) = \Delta {r_m}\cos {\theta _m} + \sqrt {{x^2} + {{\left( {\Delta {r_m}\cos \left( {{\theta _m}} \right)} \right)}^2} - \Delta {r_m}^2} \quad .\tag{43b}\]
\end{figure*}

Substitute (43a) into (41a), the coverage probability of the moving user is derived by (44).
\begin{figure*}[!t]
\[\begin{gathered}
  {P_{m\_\operatorname{cov} er}} = \int\limits_{x > 0} {\left( {\frac{1}{\pi }\int\limits_0^\pi  {\frac{{2\pi x{\lambda _{m,{\text{BS}}}}{\omega _m}\left( {x,{\theta _m}} \right){e^{ - \pi {\lambda _{m,{\text{BS}}}}{{\left( {{\omega _m}\left( {x,{\theta _m}} \right)} \right)}^2}}}}}{{{\omega _m}\left( {x,{\theta _m}} \right) - \Delta {r_m}\cos {\theta _m}}}d{\theta _m}} } \right)}  \hfill \\
  \quad \quad \quad \; \cdot \;\int\limits_{ - \infty }^\infty  {\frac{{{e^{ - 2\pi {\sigma ^2}sj}}{e^{ - 2\pi x\int\limits_x^\infty  {\left( {1 - \frac{{{\lambda _h}}}{{{\lambda _h} + 2\pi s{v^{ - \alpha }}j}}} \right)vdv} }}\left( {\frac{{{\lambda _h}}}{{{\lambda _h} - 2\pi {{\left( {{\gamma _0}x} \right)}^{ - 1}}sj}} - 1} \right)}}{{2\pi sj}}} dsdx\quad  \hfill \\
\end{gathered}. \tag{44}\]
\end{figure*}

\section{Numerical Results and Discussions}
Based on proposed user mobility probabilities and coverage probabilities in Section III and section IV, the effect of various system parameters on the user mobility probabilities and coverage probabilities have been analyzed and compared by numerical simulation in this section. In what follows, some default parameters are configured as: the area of the finite plane $\mathbb{R}_t^2$ is ${S_t} = 10{\text{k}}{{\text{m}}^2}$ , the area of the community $\mathbb{R}_c^2$ is ${S_c} = 1{\text{k}}{{\text{m}}^2}$, densities of small cell BSs inside and outside the community are ${\lambda _{c,BS}} = 20$ per square kilometers and ${\lambda _{s,BS}} = 5$ per square kilometers, respectively;  the parameters of IMM are $\rho  = 1$ ,  ${\beta _c} = 0.5$ and ${\beta _s} = 1.5$ ; the transmission power of small cell BS is ${P_t} = 0.1{\text{W}}$ , the parameter of Rayleigh fading channel is ${\lambda _h} = 1$ , the path loss exponent $\alpha  = 4$ and the velocity of user is $\overline v  = 5{\text{m/s}}$ , the user moving time is $\Delta {t_m} = 10{\text{s}}$ .

Fig. 2 shows the user pause probability with respect to the average user velocity and the area ratio of the community and the finite plane ${S_c}/{S_t}$  based on IMM and RWP (random waypoint model) \cite{Le06,Yoon06,Campos04}. When the area ratio of the community and the finite plane ${S_c}/{S_t}$  is fixed, the user pause probability ${\pi _{pause}}$  increases with the increase of the average user velocity. Considering IMM and the average user velocity is fixed, the user pause probability first decreases with the increase of ${S_c}/{S_t}$  until the value of ${S_c}/{S_t}$  reaches at $5\%$, and then the user pause probability increases with the increase of ${S_c}/{S_t}$ . Hence, the user pause probability achieves a minimum when the area ratio of the community and the finite plane ${S_c}/{S_t}$  is configured as $5\%$. This result provides a basic guideline to evaluate the user pause probability in a region with different ratio of hot sports. Furthermore, the deployment of small cell BSs can be considered for the user pause probability, especial in the urban region with different ratio of hot sports. Considering RWP and the average user velocity is fixed, the user pause probability is a constant without respect to the area ratio of the community and the finite plane ${S_c}/{S_t}$ . Therefore, the user pause probability based on IMM considers the human clustering behavior, e.g. people would like to stay at the community, in 5G small cell networks.
\begin{figure}
\vspace{0.1in}
\centerline{\includegraphics[width=8cm,draft=false]{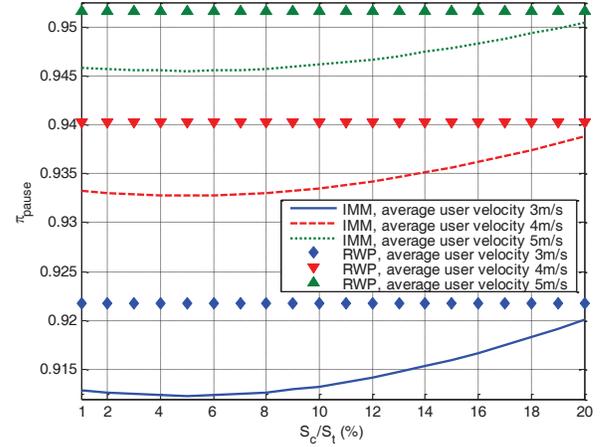}}
\caption{\small User pause probability with respect to the average user velocity and the area ratio of the community and the finite plane ${S_c}/{S_t}$ .}
\end{figure}

Fig. 3 illustrates the user arrival probability  ${\pi _{c,in}}$ with respect to the average user velocity and the area ratio of the community and the finite plane ${S_c}/{S_t}$ . When the IMM is considered in Fig. 3, numerical results are explained as follows. When the area ratio ${S_c}/{S_t}$  is fixed, the user arrival probability ${\pi _{c,in}}$  decreases with the increase of the average user velocity. When the average user velocity is fixed, the user arrival probability ${\pi _{c,in}}$  increases with the increase of the area ratio ${S_c}/{S_t}$ . When the RWP is considered in Fig. 3, the user arrival probability ${\pi _{c,in}}$  is independent with the average user velocity and increases with the increase of the area ratio ${S_c}/{S_t}$ . Compared with curves in Fig. 3, the user arrival probability based on IMM is larger than the user arrival probability based on RWP. This result implies that people would like to stay at the community in 5G small cell networks considering the human tendency behavior.

\begin{figure}
\vspace{0.1in}
\centerline{\includegraphics[width=8cm,draft=false]{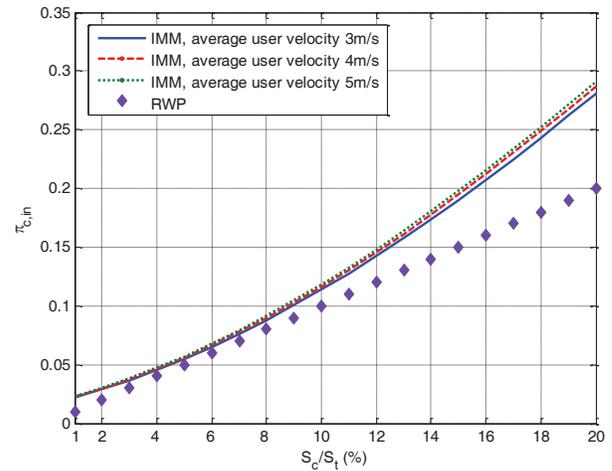}}
\caption{\small User arrival probability ${\pi _{c,in}}$  with respect to the average user velocity and the area ratio of the community and the finite plane ${S_c}/{S_t}$ .}
\end{figure}

In Fig. 4, the impact of the SINR threshold ${\gamma _0}$  and the density of small cell BSs inside and outside the community on the number of available small cell BSs ${N_{cover}}$  for a static user is evaluated. When the density of small cell BSs outside the community is fixed as ${\lambda _{s,BS}} = 5$  , the number of available small cell BSs  ${N_{cover}}$ for a static user with respect to the SINR threshold ${\gamma _0}$  and the density of small cell BSs ${\lambda _{c,BS}}$  inside the community is depicted in Fig. 4(a). When the density of small cell BSs ${\lambda _{c,BS}}$  is fixed, the number of available small cell BSs for a static user decreases with the increase of the SINR threshold. When the SINR threshold is fixed, the number of available small cell BSs for a static user increases with the increase of the density of small cell BSs inside the community. When the density of small cell BSs inside the community is fixed as ${\lambda _{c,BS}} = 20$ , the number of available small cell BSs ${N_{cover}}$  for a static user with respect to the SINR threshold ${\gamma _0}$  and the density of small cell BSs ${\lambda _{s,BS}}$  outside the community is described in Fig. 4(b). When the SINR threshold is fixed, the number of available small cell BSs for a static user increases with the increase of the density of small cell BSs outside the community. Compared with curves in Fig. 4(a) and Fig. 4(b), the number of available small cell BSs based on IMM is less than the number of available small cell BSs based on RWP. This result implies that the number of small cell BSs used for cooperative communications is overestimate for small cell networks when RWP is used for the user mobility model.

\begin{figure}
\vspace{0.1in}
\centerline{\includegraphics[width=8cm,draft=false]{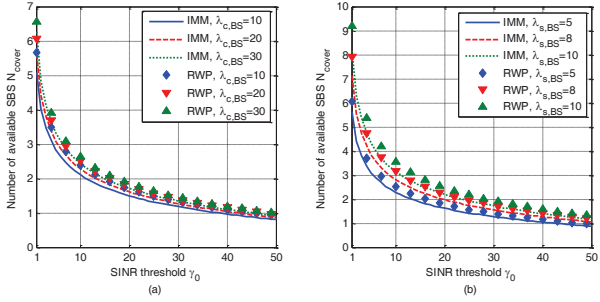}}
\caption{\small  Number of available small cell BSs ${N_{cover}}$  with respect to the SINR threshold ${\gamma _0}$  and the density of small cell BSs inside and outside the community. }
\end{figure}

In Fig. 5, the effect of the SINR threshold ${\gamma _0}$  and the density of small cell BSs ${\lambda _{s,BS}}$  on the coverage probability of a small cell BS ${P_{c\_cover}}$  is investigated inside the community. When the density of small cell BSs ${\lambda _{c,BS}}$  is fixed, the coverage probability of a small cell BS ${P_{c\_cover}}$  decreases with the increase of the SINR threshold. When the SINR threshold ${\gamma _0}$  is fixed, the coverage probability of a small cell BS ${P_{c\_cover}}$  decreases with the increase of the density of small cell BSs inside the community.

\begin{figure}
\vspace{0.1in}
\centerline{\includegraphics[width=8cm,draft=false]{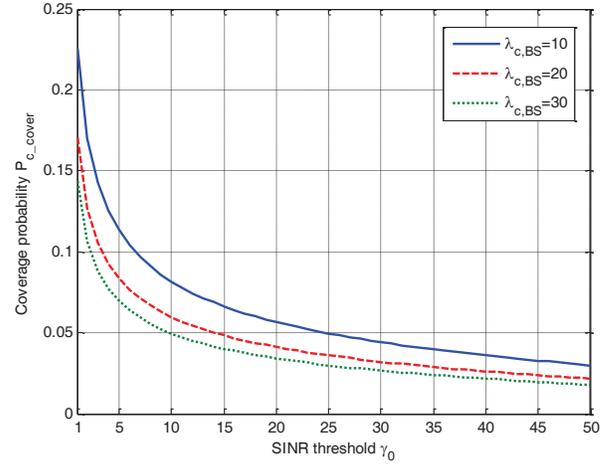}}
\caption{\small  Coverage probability of a small cell with respect to the SINR threshold and the density of small cell BSs inside the community.}
\end{figure}

In Fig. 6, the impact of the SINR threshold ${\gamma _0}$  and the density of small cell BSs  ${\lambda _{s,BS}}$ on the coverage probability of a small cell BS ${P_{s\_cover}}$  is analyzed outside the community. When the density of small cell BSs ${\lambda _{s,BS}}$  is fixed, the coverage probability of a small cell BS ${P_{s\_cover}}$  decreases with the increase of the SINR threshold. When the SINR threshold ${\gamma _0}$  is fixed, the coverage probability of a small cell BS ${P_{s\_cover}}$  decreases with the increase of the density of small cell BSs outside the community.

\begin{figure}
\vspace{0.1in}
\centerline{\includegraphics[width=8cm,draft=false]{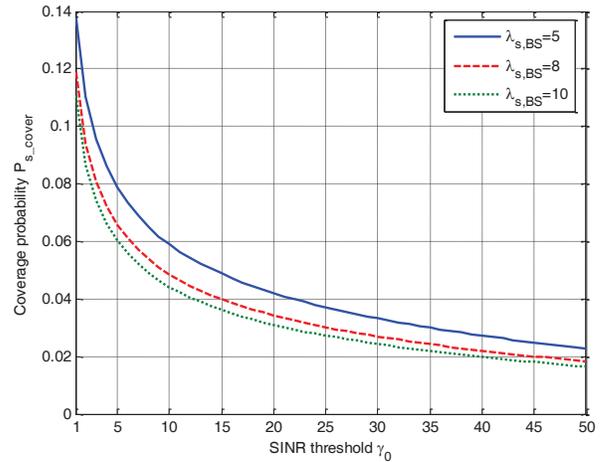}}
\caption{\small Coverage probability of a small cell BS with respect to the SINR threshold and the density of small cell BSs outside the community. }
\end{figure}

Finally, Fig. 7 evaluates the coverage probability of a macro cell BS ${P_{m\_cover}}$  with respect to the SINR threshold and the average user velocity. When the SINR threshold is fixed, the coverage probability of a macro cell BS decreases with the increase of the average user velocity. When the average user velocity is fixed, the coverage probability of a macro cell BS decreases with the increase of the SINR threshold.

\begin{figure}
\vspace{0.1in}
\centerline{\includegraphics[width=8cm,draft=false]{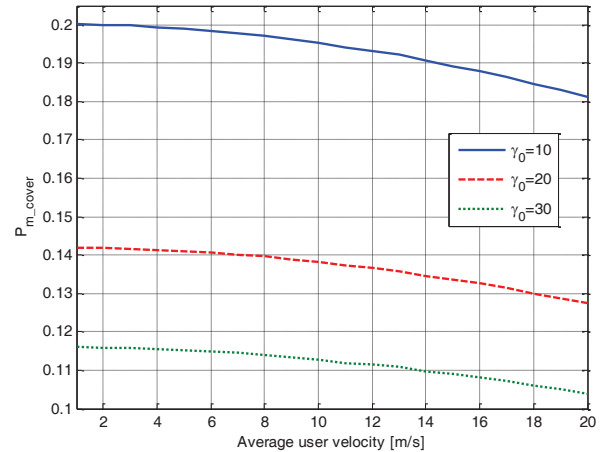}}
\caption{\small  Coverage probability of a macro cell BS ${P_{m\_cover}}$  with respect to the SINR threshold and the average user velocity. }
\end{figure}

\section{Conclusion}
Based on the IMM, the impact of user mobility performance on 5G small cell networks is evaluated in this paper. Considering the hot sports scenario in the urban region, a system model with the community is proposed for 5G small cell networks based on IMM. Moreover, the user pause probability, the user arrival and departure probabilities are derived for evaluating the user mobility performance in 5G small cell networks considering the hot sport scenario. Furthermore, the coverage probabilities of a small cell BS inside and outside the community are derived for the static user in 5G small cell networks, respectively. Besides, the coverage probability of a macro cell BS is derived for the moving user in 5G small cell networks. These results are very useful to investigate issues such as How to deploy small cell BSs to improve the network performance, such as the BS coverage probability considering the human tendency and clustering behaviors in real world.

As we mentioned in Section I, existing studies have not investigated the impact of human tendency and clustering behaviors on 5G small cell networks. Based on IMM, this paper is the first work that evaluates the user mobility performance for 5G small cell Networks considering human tendency and clustering behaviors.

\section*{Appendix A}
To simplify the derivation, terms in ${t_{c,in}}\left( n \right)$  are denoted by (45)
\begin{figure*}[!t]
\[\left\{ \begin{gathered}
  {B_1} = \sum\limits_{i = 1}^{{n_{i,i}}} {\left[ {\frac{{\left\| {L\left( I \right) - L\left( {k,k \in \left[ {0,I - 1} \right]} \right)} \right\|}}{{\overline v }}{\mathbf{1}}\left( \mathbb{A} \right) + \frac{{\left\| {L\left( I \right) - L\left( {k,k \in \left[ {0,I - 1} \right]} \right)} \right\|}}{{\overline v }}{\mathbf{1}}\left( \mathbb{B} \right)} \right]}  \hfill \\
  {B_2} = \sum\limits_{i = 1}^{{n_{o,i}}} {\left[ {\frac{{\left\| {L\left( I \right) - L\left( {k,k \in \left[ {0,I - 1} \right]} \right)} \right\|}}{{\overline v }}{\mathbf{1}}\left( \mathbb{A} \right) + \frac{{\left\| {L\left( I \right) - L\left( {k,k \in \left[ {0,I - 1} \right]} \right)} \right\|}}{{\overline v }}{\mathbf{1}}\left( \mathbb{B} \right)} \right]}  \hfill \\
  {B_3} = \sum\limits_{i = 1}^{{n_{i,o}}} {\left[ {\frac{{\left\| {L\left( I \right) - L\left( {k,k \in \left[ {0,I - 1} \right]} \right)} \right\|}}{{\overline v }}{\mathbf{1}}\left( \mathbb{A} \right) + \frac{{\left\| {L\left( I \right) - L\left( {k,k \in \left[ {0,I - 1} \right]} \right)} \right\|}}{{\overline v }}{\mathbf{1}}\left( \mathbb{B} \right)} \right]}  \hfill \\
  {B_4} = {n_{c,in}}\Delta {t_c} \hfill \\
\end{gathered}  \right.\quad .\tag{45}\]
\end{figure*}

Based on (45), (10) can be simply expressed by (46)
\begin{figure*}[!t]
\[\begin{gathered}
\begin{aligned}
  {\pi _{c,in}}& = \mathop {\lim }\limits_{n \to \infty } \mathbb{E}\left[ {\frac{{{B_1} + {B_2} + {B_3} + {B_4}}}{{\overline v t\left( n \right)}}} \right] \hfill \\
  &= \mathop {\lim }\limits_{n \to \infty } \left[ {\mathbb{E}\left[ {\frac{{{B_1}}}{{\overline v t\left( n \right)}}} \right] + \mathbb{E}\left[ {\frac{{{B_2}}}{{\overline v t\left( n \right)}}} \right] + \mathbb{E}\left[ {\frac{{{B_3}}}{{\overline v t\left( n \right)}}} \right] + \mathbb{E}\left[ {\frac{{{B_4}}}{{\overline v t\left( n \right)}}} \right]} \right] \hfill \\
  &= \mathop {\lim }\limits_{n \to \infty } \mathbb{E}\left[ {\frac{{{B_1}}}{{\bar vt\left( n \right)}}} \right] + \mathop {\lim }\limits_{n \to \infty } \mathbb{E}\left[ {\frac{{{B_2}}}{{\bar vt\left( n \right)}}} \right] + \mathop {\lim }\limits_{n \to \infty } \mathbb{E}\left[ {\frac{{{B_3}}}{{\bar vt\left( n \right)}}} \right] + \mathop {\lim }\limits_{n \to \infty } \mathbb{E}\left[ {\frac{{{B_4}}}{{\bar vt\left( n \right)}}} \right] \hfill \\
  \end{aligned}
\end{gathered}. \tag{46}\]
\end{figure*}

Without loss of generality, we first derive the first term of (46).

Based on (10), we can get a result that $\mathop {\lim }\limits_{n \to \infty } t\left( n \right) \to \infty $ . Assume that the number of exploring new locations is ${m_e}$  and the number of returning old locations is ${m_r}$  in the moving number of ${n_{i,i}}$ , respectively. The event $\mathbb{F}$  is that the user $K - th$  jump explores a new location and departing and arriving locations of the user $K - th$  jump are inside the community. Therefore, the following limitation is derived by
\begin{figure*}[!t]
\[\mathop {\lim }\limits_{n \to \infty } \mathbb{E}\left[ {\frac{{{m_e}}}{{t\left( n \right)}}} \right] = \mathop {\lim }\limits_{n \to \infty } \mathbb{E}\left[ {\frac{{\sum\limits_{k = 1}^n {\rho S{{\left( k \right)}^{ - \gamma }}{\mathbf{1}}\left( \mathbb{F} \right)} }}{{t\left( n \right)}}} \right] = \mathbb{E}\left[ {\mathop {\lim }\limits_{n \to \infty } \frac{{\sum\limits_{k = 1}^n {\rho S{{\left( k \right)}^{ - \gamma }}{\mathbf{1}}\left( \mathbb{F} \right)} }}{{t\left( n \right)}}} \right]\quad .\tag{47}\]
\end{figure*}
Let ${\Theta _1} = \left\{ {\rho {z^{ - \gamma }},z \in {\mathbb{Z}^ + }} \right\}$  is a series, where ${\mathbb{Z}^ + }$  is a positive integer set. It is obviously that {\footnotesize {${\Theta _2} = \left\{ {\rho S{{\left( k \right)}^{ - \gamma }}{\mathbf{1}}\left( \mathbb{F} \right),k \in {\mathbb{Z}^ + }} \right\}$}} is a subseries of ${\Theta _1}$ . Based on the result in \cite{James11},  $\mathop {\lim }\limits_{n \to \infty } \sum\limits_{z = 1}^n {\rho {z^{ - \gamma }}} $ converges to a limited value when $\gamma  > 1$ . When $\gamma  > 1$ , it is also derived that the sum of the subseries of ${\Theta _1}$ , i.e., $\mathop {\lim }\limits_{n \to \infty } \sum\limits_{k = 1}^n {\rho S{{\left( k \right)}^{ - \gamma }}{\mathbf{1}}\left( \mathbb{F} \right)} $  also converges to a limited value, which is denoted as $\delta $ . Moreover, (47) is derived by
\begin{figure*}[!t]
\[\mathop {\lim }\limits_{n \to \infty } \mathbb{E}\left[ {\frac{{{m_e}}}{{t\left( n \right)}}} \right] = \mathbb{E}\left[ {\mathop {\lim }\limits_{n \to \infty } \frac{{\sum\limits_{k = 1}^n {\rho S{{\left( k \right)}^{ - \gamma }}{\mathbf{1}}\left( \mathbb{F} \right)} }}{{t\left( n \right)}}} \right] = \mathbb{E}\left[ {\mathop {\lim }\limits_{n \to \infty } \frac{\delta }{{t\left( n \right)}}} \right] = 0\quad .\tag{48}\]
\end{figure*}
Furthermore, the limitation is derived by
\begin{figure*}[!t]
{\footnotesize {
\[\begin{gathered}
\begin{aligned}
  \mathop {\lim }\limits_{n \to \infty } \mathbb{E}\left[ {\frac{{{B_1}}}{{t\left( n \right)}}} \right] &= \mathop {\lim }\limits_{n \to \infty } \mathbb{E}\left[ {\sum\limits_{i = 1}^{{n_{i,i}}} {\left[ {\frac{{\left\| {L\left( I \right) - L\left( {k,k \in \left[ {0,I - 1} \right]} \right)} \right\|}}{{\overline v }}{\mathbf{1}}\left( \mathbb{A} \right) + \frac{{\left\| {L\left( I \right) - L\left( {k,k \in \left[ {0,I - 1} \right]} \right)} \right\|}}{{\overline v }}{\mathbf{1}}\left( \mathbb{B} \right)} \right]} /t\left( n \right)} \right] \hfill \\
  &= \mathop {\lim }\limits_{n \to \infty } \mathbb{E}\left[ {\frac{{\left[ {{m_e}\frac{{\left\| {L\left( I \right) - L\left( {k,k \in \left[ {0,I - 1} \right]} \right)} \right\|}}{{\overline v }} + {m_r}\frac{{\left\| {L\left( I \right) - L\left( {k,k \in \left[ {0,I - 1} \right]} \right)} \right\|}}{{\overline v }}} \right]}}{{t\left( n \right)}}} \right] \hfill \\
  &= \mathop {\lim }\limits_{n \to \infty } \mathbb{E}\left[ {\frac{{{m_e}\frac{{\left\| {L\left( I \right) - L\left( {k,k \in \left[ {0,I - 1} \right]} \right)} \right\|}}{{\overline v }}}}{{t\left( n \right)}} + \frac{{{m_r}\frac{{\left\| {L\left( I \right) - L\left( {k,k \in \left[ {0,I - 1} \right]} \right)} \right\|}}{{\overline v }}}}{{t\left( n \right)}}} \right] \hfill \\
  &= \mathop {\lim }\limits_{n \to \infty } \mathbb{E}\left[ {\frac{{{m_r}\frac{{\left\| {L\left( I \right) - L\left( {k,k \in \left[ {0,I - 1} \right]} \right)} \right\|}}{{\overline v }}}}{{t\left( n \right)}}} \right] \hfill \\
  &= \mathop {\lim }\limits_{n \to \infty } \mathbb{E}\left[ {\frac{{\left( {{n_{i,i}} - \delta } \right)\left\| {L\left( I \right) - L\left( {k,k \in \left[ {0,I - 1} \right]} \right)} \right\|}}{{\overline v t\left( n \right)}}} \right] \hfill \\
  &= \mathop {\lim }\limits_{n \to \infty } \mathbb{E}\left[ {\frac{{{n_{i,i}}\left\| {L\left( I \right) - L\left( {k,k \in \left[ {0,I - 1} \right]} \right)} \right\|}}{{\overline v t\left( n \right)}} - \frac{{\delta \left\| {L\left( I \right) - L\left( {k,k \in \left[ {0,I - 1} \right]} \right)} \right\|}}{{\overline v t\left( n \right)}}} \right] \hfill \\
  &= \mathop {\lim }\limits_{n \to \infty } \mathbb{E}\left[ {\frac{{{n_{i,i}}\left\| {L\left( I \right) - L\left( {k,k \in \left[ {0,I - 1} \right]} \right)} \right\|}}{{\overline v t\left( n \right)}}} \right]\quad  \hfill \\
\end{aligned}
\end{gathered}. \normalsize{\tag{49}}\]
}}
\end{figure*}
Utilizing the same method, other terms in ${\pi _{c,in}}$  are derived by
\begin{figure*}[!t]
\[\left\{ \begin{gathered}
  \mathop {\lim }\limits_{n \to \infty } \mathbb{E}\left[ {\frac{{{B_2}}}{{t\left( n \right)}}} \right] = \mathop {\lim }\limits_{n \to \infty } \mathbb{E}\left[ {\frac{{{n_{o,i}}\left\| {L\left( I \right) - L\left( {k,k \in \left[ {0,I - 1} \right]} \right)} \right\|}}{{\overline v t\left( n \right)}}} \right] \hfill \\
  \mathop {\lim }\limits_{n \to \infty } \mathbb{E}\left[ {\frac{{{B_3}}}{{t\left( n \right)}}} \right] = \mathop {\lim }\limits_{n \to \infty } \mathbb{E}\left[ {\frac{{{n_{i,o}}\left\| {L\left( I \right) - L\left( {k,k \in \left[ {0,I - 1} \right]} \right)} \right\|}}{{\overline v t\left( n \right)}}} \right] \hfill \\
  \mathop {\lim }\limits_{n \to \infty } \mathbb{E}\left[ {\frac{{{B_4}}}{{t\left( n \right)}}} \right] = \mathop {\lim }\limits_{n \to \infty } \mathbb{E}\left[ {\frac{{\left( {{n_{o,i}} + {n_{i,i}}} \right)\Delta {t_c}}}{{\overline v t\left( n \right)}}} \right] \hfill \\
\end{gathered}  \right.\quad ,\tag{50}\]
\end{figure*}
Substitute (49) and (50) into (46), (11) can be derived.


\begin{IEEEbiography}[{\includegraphics[width=1in,height=1.25in,clip,keepaspectratio]{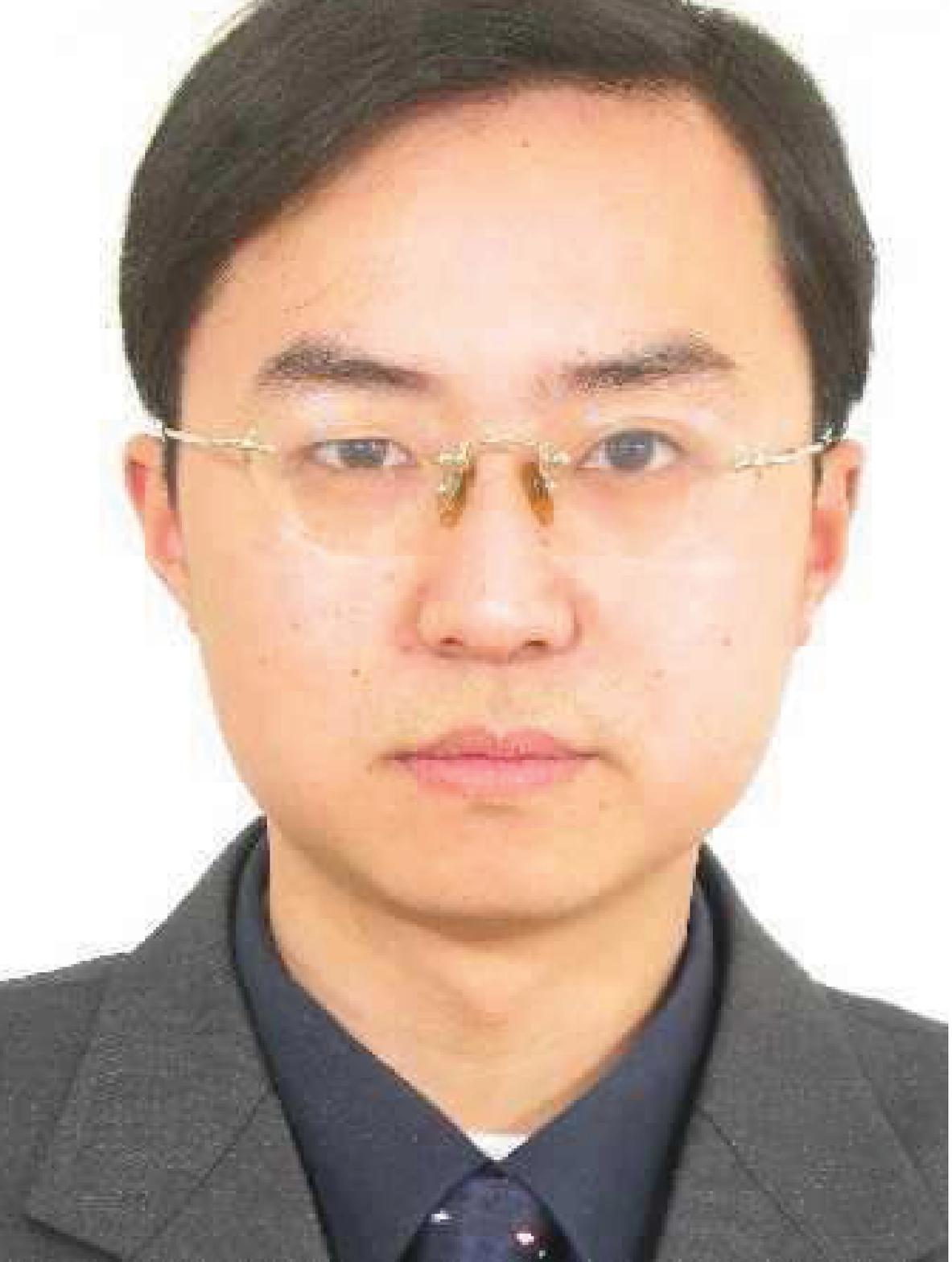}}]{Xiaohu~Ge}
(M'09-SM'11) is currently a Professor with the School of Electronic Information and Communications at Huazhong University of Science and Technology (HUST), China. He received his PhD degree in Communication and Information Engineering from HUST in 2003. He has worked at HUST since Nov. 2005. Prior to that, he worked as a researcher at Ajou University (Korea) and Politecnico Di Torino (Italy) from Jan. 2004 to Oct. 2005. He was a visiting researcher at Heriot-Watt University, Edinburgh, UK from June to August 2010. His research interests are in the area of mobile communications, traffic modeling in wireless networks, green communications, and interference modeling in wireless communications. He has published about 90 papers in refereed journals and conference proceedings and has been granted about 15 patents in China. He received the Best Paper Awards from IEEE Globecom 2010. He is leading several projects funded by NSFC, China MOST, and industries. He is taking part in several international joint projects, such as the EU FP7-PEOPLE-IRSES: project acronym S2EuNet (grant no. 247083), project acronym WiNDOW (grant no. 318992) and project acronym CROWN (grant no. 610524).

Dr. Ge is a Senior Member of the China Institute of Communications and a member of the National Natural Science Foundation of China and the Chinese Ministry of Science and Technology Peer Review College. He has been actively involved in organizing more the ten international conferences since 2005. He served as the general Chair for the 2015 IEEE International Conference on Green Computing and Communications (IEEE GreenCom). He serves as an Associate Editor for the \textit{IEEE ACCESS}, \textit{Wireless Communications and Mobile Computing Journal (Wiley)} and \textit{the International Journal of Communication Systems (Wiley)}, etc.
Moreover, he served as the guest editor for \textit{IEEE Communications Magazine} Special Issue on 5G Wireless Communication Systems.
\end{IEEEbiography}
\vspace{-6 mm}

\begin{IEEEbiography}[{\includegraphics[width=1in,height=1.25in,clip,keepaspectratio]{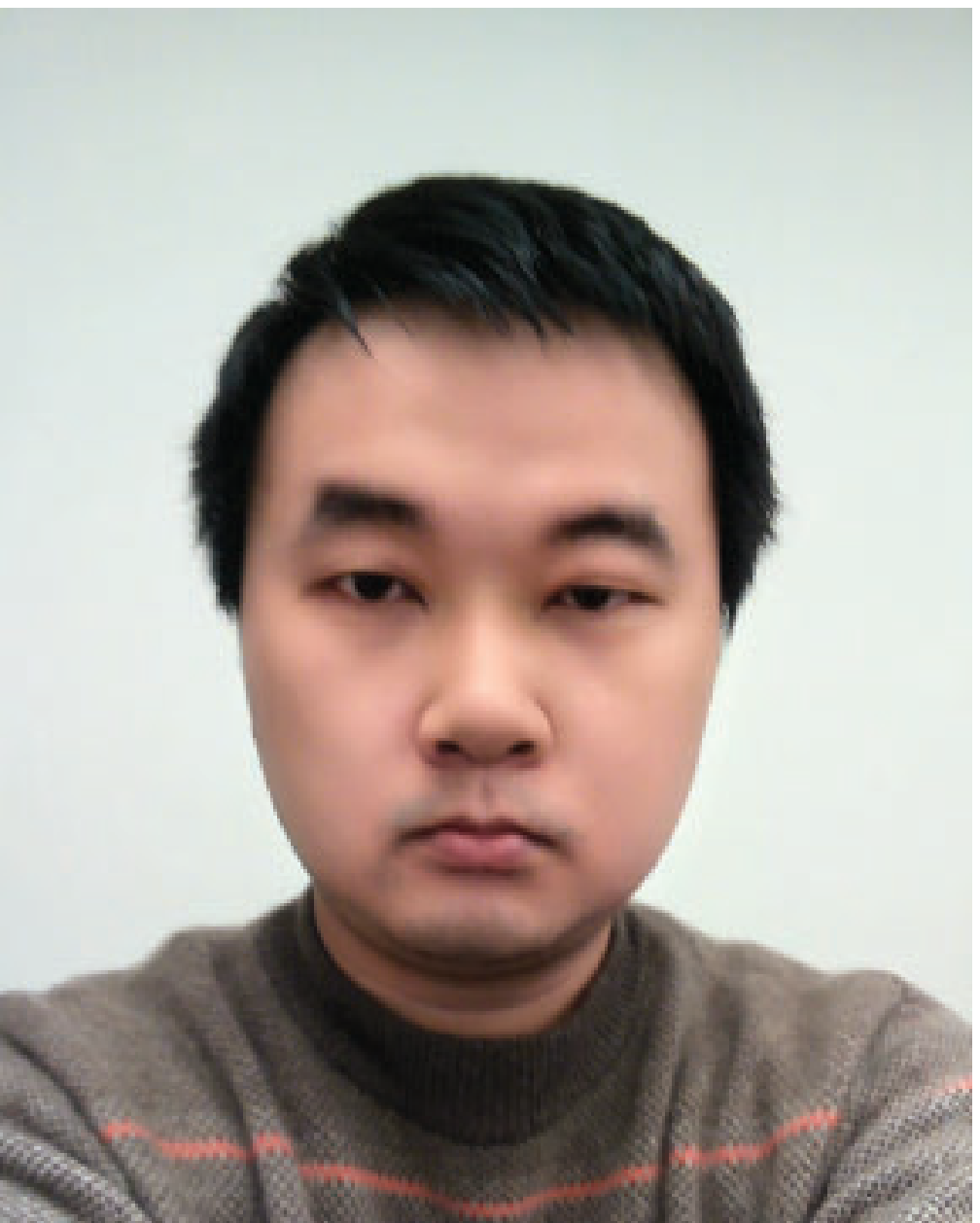}}]{Junliang~Ye}
recieved the B.Sc. Degree in communication engineering from China University of Geosciences, Wuhan, P.R China, in 2011, and he is currently a Ph.D student in communication and information system in Huazhong University of Science and Technology, Wuhan, P.R China.

His research interests include heterogeneous networks, stochastic geometry, mobility based access models of cellular networks and next generation wireless communication.
\end{IEEEbiography}
\vspace{-6 mm}

\begin{IEEEbiography}[{\includegraphics[width=1in,height=1.25in,clip,keepaspectratio]{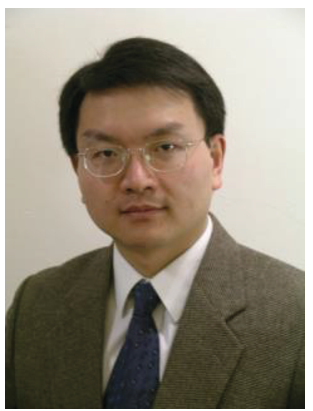}}]{Yang~Yang}
(SM'10) received the BEng and MEng degrees in Radio Engineering from Southeast University, Nanjing, P. R. China, in 1996 and 1999, respectively; and the PhD degree in Information Engineering from The Chinese University of Hong Kong in 2002.

Dr. Yang Yang is currently a Professor with the School of Information Science and Technology, ShanghaiTech University, and the Director of Shanghai Research Center for Wireless Communications (WiCO). Prior to that, he has served Shanghai Institute of Microsystem and Information Technology (SIMIT), Chinese Academy of Sciences, as a Professor; the Department of Electronic and Electrical Engineering at University College London (UCL), United Kingdom, as a Senior Lecturer; the Department of Electronic and Computer Engineering at Brunel University, United Kingdom, as a Lecturer; and the Department of Information Engineering at The Chinese University of Hong Kong as an Assistant Professor. His research interests include wireless ad hoc and sensor networks, wireless mesh networks, next generation mobile cellular systems, intelligent transport systems, and wireless testbed development and practical experiments.

Dr. Yang Yang has co-edited a book on heterogeneous celluar networks (2013, Cambridge University Press) and co-authored more than 100 technical papers. He has been serving in the organization teams of about 50 international conferences, e.g. a co-chair of Ad-hoc and Sensor Networking Symposium at IEEE ICC¡¯15, a co-chair of Communication and Information System Security Symposium at IEEE Globecom¡¯15.
\end{IEEEbiography}

\vspace{-6 mm}

\begin{IEEEbiography}[{\includegraphics[width=1in,height=1.25in,clip,keepaspectratio]{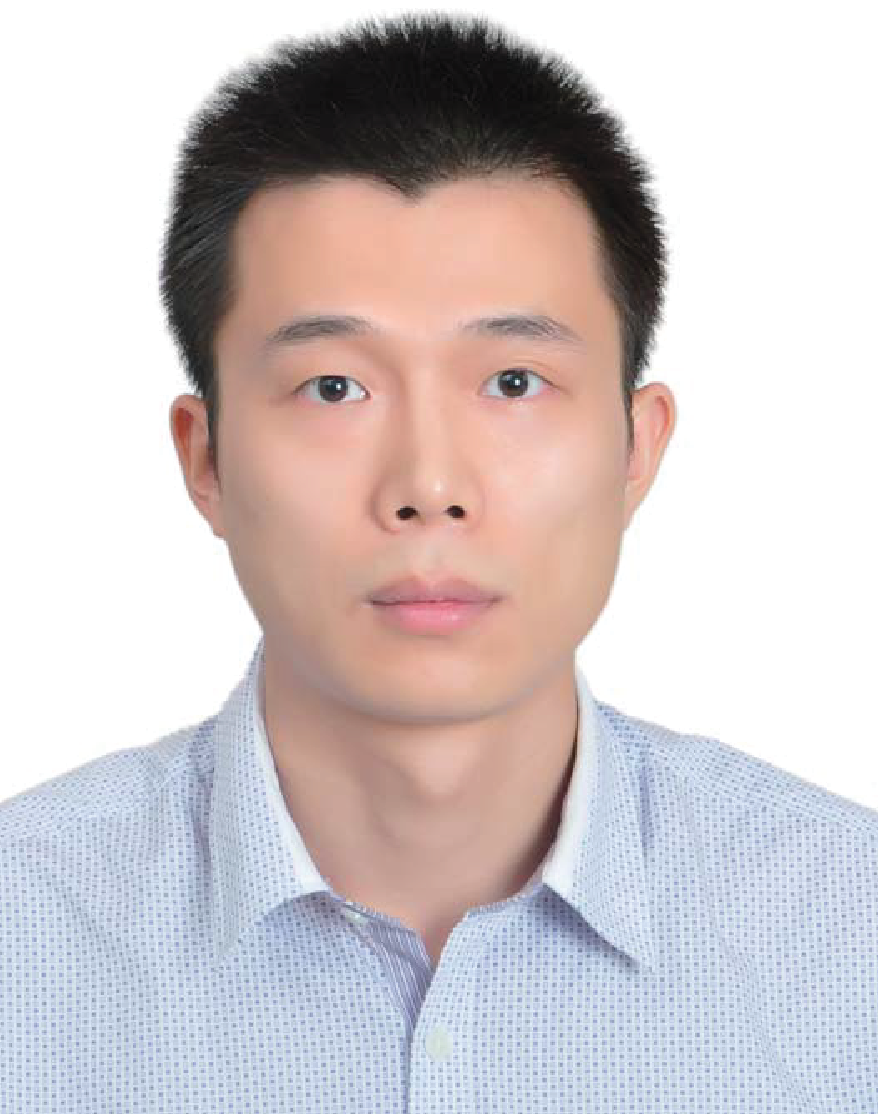}}]{Qiang~Li}
Qiang Li received the B.Eng. degree in communication engineering from the University of Electronic Science and Technology of China (UESTC), Chengdu, China, in 2007 and the Ph.D. degree in electrical and electronic engineering from Nanyang Technological University (NTU), Singapore, in 2011. From 2011 to 2013, he was a Research Fellow with Nanyang Technological University. Since 2013, he has been an Associate Professor with Huazhong University of Science and Technology (HUST), Wuhan, China. His current research interests include future broadband wireless networks, cooperative communications, wireless power transfer, cognitive radio networks.
\end{IEEEbiography}

\end{document}